\documentclass[journal=jpcbfk, manuscript=article]{achemso}

\usepackage[version=3]{mhchem} % Formula subscripts using \ce{}
\usepackage {amsmath}

\usepackage {color}
\usepackage{soul}
\usepackage{comment}
\usepackage{siunitx}

\author{Peiyuan Gao}
\email{Peiyuan.Gao@pnnl.gov}
\author{Alexandre M. Tartakovsky}
\email{Alexandre.Tartakovsky@pnnl.gov}
\affiliation
{Pacific Northwest National Laboratory, Richland, Washington 99352, United States}

\title
{MARTINI-based Coarse-grained Model for Poly ($\alpha$-peptoid)s}

\begin{document}

\begin{abstract}
In this paper, we present a new coarse-grained (CG) model for
poly ($\alpha$-peptoid)s that is compatible with the MARTINI CG FF. 
In the proposed model, CG poly ($\alpha$-peptoid) is composed by a CG backbone (here we select polysarcosine as the backbone) and side chain beads. The CG model of the backbone (polysarcosine) in a solvent is first developed
and then extended to poly ($\alpha$-peptoid)s with different side groups that can be obtained from MARTINI FF. We demonstrate that our CG model has good transferability. For example, the CG potentials for polysarcosine can be transferred to predict hydration free energy of other peptoids. Also, the CG polypeptoid model accurately predicts the radius of gyration over a wide range of chain lengths and the solvation free energy for relatively short peptoid molecules in good solvents. We use the CG model to study sequenced diblock polypeptoid in binary solvent
mixtures and compare the results with the experimentally observed coil-globule transition.  

%\st{Transferability of the nonbonded potential and bond and angle potentials of the CG backbone to other
%poly $\alpha$-peptoids in good solvent is demonstrated. We also show that in good solvent, the CG torsion potential is transferable, when only van der Waals forces exist between the side groups for the molecule.  when other special interactions as hydrogen bonding between side groups are present, the torsion potential is not transferable and must be computed for each specific peptoid
%To further validate the CG
%FF, we model sequenced diblock poly $\alpha$-peptoid in binary solvent
%mixtures and compare results with an experimentally-observed coil-globule transition.}
\end{abstract}

\section{Introduction}
Peptoids are artificial polymers designed to mimic functions of
naturally-occurring peptides. In peptides, side chains are appended to
the {$\alpha$}-carbon, while in peptoids, side chains are attached to
nitrogen atoms and form repeat units of N-substituted glycine
molecules\cite{RN70}. The lack of both backbone chirality and
backbone hydrogen bond donors in peptoids results in a variety of
secondary structures. Peptoid biomimetic structure and well controllable
molecular sequence\cite{RN275} have been shown to benefit
applications ranging from biomedicine to material synthesis.\cite{RN343} For example, peptoids have been used in biomineralization\cite{RN386,RN685},
antifouling\cite{RN277}, hydrate inhibitors\cite{RN125} and
biorecognition sensors\cite{RN182}.
Similar to amino acids, peptoids can be classified as $\alpha$-peptoids,
$\beta$-peptoids, and $\gamma$-peptoids according to the N-substituted group
position. Among them, oligomeric $\alpha$-peptoids have been extensively
investigated as peptidomimetic surrogates for medical applications.
%Despite lacking the extensive backbone hydrogen bonding network of $\alpha$-peptide, 
Poly($\alpha$-peptoid)s 
%\comment{Should the title of the article also be punctuated this way, poly($\alpha$-peptoid)s?} 
can fold into well-defined secondary
structures (e.g., helices) dictated by the steric and electronic
properties of the side chains. The simplest example of such structures
is polysarcosine, based on the natural, non-toxic amino acid
sarcosine (N-methyl glycine).\cite{RN664} In the past,
polysarcosine was mostly considered in the context of synthetic
polypeptides. Recently, polysarcosine have been rediscovered as a
biocompatible and degradable polymer and employed in a number of drug
delivery systems as micelles,\cite{RN387}
polymersomes,\cite{RN388} protein
conjugates,\cite{RN389} and nanoparticles.\cite{RN348}
Furthermore, polysarcosine-based block copolymers, especially
polysarcosine-co-polypeptides or co-polypeptoids, bear enormous
potential to create body-compatible materials enabling the synthesis of
carrier systems completely based on endogenous amino
acids.\cite{RN122} 

The first polysarcosine block copolymers were
reported by Gallot\cite{RN394} and Kimura\cite{RN396}.
Barz and co-workers further advanced synthesis methods and produced
several novel functional block copolypept(o)ides.\cite{RN378,RN393,RN286,RN392}.
Despite significant progress in understanding peptoid block polymers,
many challenges still remain. For example, the phase space of different side chains
and conformations of polypeptoids have hardly been
explored. Since laboratory experiments are difficult to perform,
molecular simulations have become a popular tool for design-screening and discovery of new monomer and sequences. For
example, Park and Szleifer used atomic molecular dynamics (MD)
simulation to demonstrate the ability of polysarcosine and
N-methoxyethyl glycine oligomers to act as anti-fouling agents when
end-grafted to surfaces.\cite{RN40} 
%\comment{I think that ``nonfouling'' might not be the right word... the cited article might mean to say ``anti-fouling.'' The paper \textit{they} cite consistently says anti-fouling.}
Whitelam's atomic MD
simulations discovered a new secondary structure, the sigma-strand, in
polypeptoid nanosheets.\cite{RN402} Baer's atomic MD study of
peptoid oligomers\cite{RN400} improved understanding of how
hydrophobic effects and ion-mediated interactions cooperate to drive
assembly and folding in polypeptoids. These examples show that atomic MD
methods can accurately describe the solvation behavior of peptoids in
solution, including local chain orientation and intermolecular
interactions at the nanometer scale. However, because of the long-range electrostatic interactions
and large relaxation times of polypeptoid solutions, atomic MD simulations are too costly to simulate the
intermediate structure and assembly dynamics of polypeptoids. Coarse-grained (CG)
models can provide an alternative to atomistic
models.\cite{RN521} In CG models, the number of degrees of
freedom in polymer repeat units is reduced  in a systematic manner by representing a group of atoms or repeat units with a CG bead such that   critical chemical information is retained to distinguish the
interactions of various functional groups in the polymer. Therefore, a CG
model can provide an in-depth picture of nanostructures and formations (e.g., a specific backbone conformation) at significantly
reduced computational cost.  
The effective interactions between CG beads are obtained by averaging
over the atomic degree of freedom. Depending on a quantity of
interest, the methods to develop CG models can be classified as
structure-based, force-based, and thermodynamics-based
methods. The CG potentials can be obtained to reproduce
microscopic (bottom-up approach) or macroscopic (top-down approach)
quantities.

Two important properties of a ``predictive'' CG model are representability and  transferability. Representability is
the ability of a CG model to predict properties, other than those used
to construct the CG model at the same thermodynamic ``state point.'' 
Transferability refers to the ability of a CG model developed for one
kind of molecules (e.g., molecule A) at one state point to predict the
same observables for the same system at different state points, or
another molecule, which include the same blocks as molecule A. Here, a
state point includes both physical conditions (e.g.,
temperature, pressure, and an external field) and chemical
environment (e.g., CG bead is a fragment of a molecule). Generally,
the change of state point influences the thermodynamics of solvents as
well as their structure. Therefore, understanding and removing
transferability-related limitations of CG models is crucial for
predictive modeling of complex systems.\cite{RN634}

Several transferable CG models for polymers have been proposed. For example, a CG model of polystyrene,  obtained
by iterative Boltzmann inversion (IBI) method, was successfully extended
to analogue poly(4-tert-butylstyrene).\cite{RN528} However,
transferability with respect to chemical conditions is more challenging
to achieve. Mantha and Yethiraj\cite{RN167} identified that the
strength of the CG nonbonded interactions between water and
polystyrenesulfonate changes the conformation of polystyrenesulfonate in
water. This indicates that the bonded interaction between CG polymer beads
could be affected by the nonbonded CG polymer--solvent interaction in a
 CG solvent.
 %, which is different from polymer melt. 
Sayar and
co-workers investigated transferability of the diphenylalanine
conformational behavior in bulk and on the interface between
water and cyclohexane.\cite{RN226} They found that a small
modification of the CG model structural and conformational properties could dramatically alter its thermodynamic properties. For a polymer
chain in solvent, they demonstrated that not only the torsion potential
but also solvation properties of the chain fragments, as well as 
interactions in the environment, must be considered to reproduce the polymer
conformational transition. Junghans and
Mukherji\cite{RN3} also indicated that the solvation thermodynamics
is dictated by the energy density within the solvation volume of the
macromolecule, the local concentration fluctuations of the solvent
components, and entropic contributions. Therefore, multiple targets
are needed to improve the transferability of a CG model for polymer in
various solvents. Several hybrid approaches with multi-targets have been
proposed to improve the transferability of CG polymer
models\cite{RN509,RN478}. For example, Sauter and
Grafmuller\cite{RN222} developed a CG model of polysaccharides
with combined structure-based and force-based CG approaches, which
improved the transferability over both concentration and degrees of
polymerization. Similarly, Abbott and Steven\cite{RN161} built the CG model
of poly(N-isopropyl acrylamide) (PNIPAM) by selecting density and
interfacial tension as target properties that successfully
captured the coil--globule transition at different temperatures. A number of CG models of proteins have been proposed in recent years \cite{RN4,RN252,RN1000,RN652,RN650}, but very few CG models of peptoids exist, and their transferability has not been
demonstrated.\cite{RN283}

In this paper, we present a new CG model for poly $\alpha$-peptoids in solution that combines the structure-based and thermodynamic-based approaches
(i.e., we select the local and global molecular structures and solvation
free energy as targets) under the
framework of MARTINI CG force field (FF).\cite{RN274}
We choose polysarcosine as our initial target molecule because of its simplicity and also because it serves a backbone of many poly $\alpha$-peptoids. Because of compatibility with 
MARTINI FF, some nonbonded interactions as interactions between beads of
different solvents and sidechain beads and solvents can be directly
borrowed from MARTINI FF.
In our CG systems, we introduce three types of beads, including
backbone, sidechain, and solvent beads.
We develop CG models of polysarcosine in four solvents: water, acetonitrile, 1-octanol, and hexane. Then, the CG poly ($\alpha$-peptoid) 
parameters is extended to other $\alpha$-peptoids as poly (N-(2-carboxyethyl) glycine) and poly N-pentyl
glycine to evaluate the transferability of the resulting CG FF. 
The performance of the CG poly
($\alpha$-peptoid) model in various solvents for different chain lengths is also tested. Finally, we apply the CG FF to a sequenced diblock polypeptoid in
binary solvent mixtures to study the coil--globule transition and validate against 
experimental results. 

\section {Methodology and simulation details}
\subsection {Atomic model}
In this work, the CG FF for poly ($\alpha$-peptoid) is built using a bottom-up
approach. Therefore, the accuracy of atomic simulations strongly
affects the CG model's accuracy. Common protein FFs, including
AMBER\cite{RN681} and OPLS-AA\cite{RN682}, cannot
accurately model peptoids.\cite{RN648,RN660} In this study, we
adopt Whitelam's atomic peptoid FF,\cite{RN17} which is based on CHARMM22 FF \cite{RN700} with optimized
torsion interactions and charge distribution. The
Whitelam peptoid FF has been validated against quantum mechanics
calculations and shown to accurately represent the local structure of
peptoids. We perform MD simulations of three
poly ($\alpha$-peptoid)s, including polysarcosine, poly (N-(2-carboxyethyl)
glycine), and poly (N-pentyl glycine). Neutral acetyl and N, N-dimethyl
amide terminal groups are added to the simulated polypeptoid chains. We calculate
the Gibbs solvation free energy (hereinafter referred to as solvation free
energy) of polysarcosine monomer in four solvents: water,
acetonitrile, 1-octanol, and hexane. The solvation free energies of poly
(N-(2-carboxyethyl) glycine) and poly (N-pentyl glycine) monomer in water
are also computed to test the transferability of CG parameters. To identify the
effect of chain length, we calculate the solvation free energy of
polysarcosine and poly (N-(2-carboxyethyl) glycine) with the length of 1, 2,
3, and 5 repeat units in acetonitrile and water. Additionally, we
compute the radius of gyration ($R_g$) of polysarcosine and poly (N-(2-carboxyethyl) glycine) in
apolar and polar solvents with the length 1, 2, 3, 5, 10, 25, 40, 50,
70, and 90 repeat units. The interaction parameters of 1-octanol,
acetonitrile, and hexane are taken from the CHARMM FF, and the TIP3P
model is used for water.\cite{RN1002} A time step of \SI{2}{fs} is used
in atomic simulations. All bonds with hydrogen atoms are
constrained using the LINCS algorithm.\cite{RN1005}
 The van der Waals forces are modeled as Lennard-Jones (LJ) force with a cutoff (set here to 0.9 nm for all atoms) and a force switch that smoothly interpolates the LJ function to zero at the distance between atoms equal to 1.2 nm.  Long-range
electrostatics is calculated using the particle-mesh Ewald summation. \cite{RN699} All modeled systems are
equilibrated using isothermal-isobaric (NPT) ensemble simulations. The
equilibrium time in these simulations is \SI{20}{ns} and the target
temperature and pressure are 298 K and 1 atm imposed with the V-rescale thermostat and
Berendsen barostat.\cite{RN635,RN636} The production simulations
are performed with Nose-Hoover thermostat\cite{RN637} and
Parrinello-Rahman barostat.\cite{RN638} The production
simulation time is \SI{2000}{ns} in the global conformation study. The
equations of motion are integrated using the velocity Verlet algorithm.
We run five independent simulations for each chain length with random
initial configurations in the global conformation study. Trajectories
are stored every 2000 steps. All MD simulations are performed using
GROMACS 5.1.2.\cite{RN639} The VMD program is used to visualize
the resulting molecular systems.\cite{RN640}
\subsection {Free energy of solvation and transfer}
To validate the atomic simulations and parameterize the CG FF, we
calculate the solvation free energy of polysarcosine, poly
(N-(2-carboxyethyl) glycine), and poly (N-pentyl glycine) in various
solvents. Several methods exist to calculate the solvation free energy,
including Bennett acceptance ratio method\cite{RN1006} (BAR),
umbrella sampling method,\cite{RN1007} and the thermodynamic
integration (TI) method.\cite{RN1008} Taddese and Carbone noted
that the BAR method give the same result as the umbrella sampling and TI
methods, while it is computationally more efficient than the TI
method.\cite{RN284} Therefore, in this study we employ the BAR
method, as implemented in GROMACS,\cite{RN639} to calculate the
solvation free energy in both the atomic and CG MD simulations. The
simulated systems include a single poly ($\alpha$-peptoid) monomer with acetyl
and N, N-dimethyl amide terminal group solvated in a simulation box
filled with: (1)~2000 water molecules, (2)~1000 1-octanol molecules, (3)~
1000 hexane molecules, and (4)~1000 acetonitrile molecules. For
polysarcosine and poly (N-(2-carboxyethyl) glycine), we also study the
effect of chain length by modeling chain with 1, 2, 3 and 5 repeat units
in acetonitrile and water.
We run ten independent simulations with random initial configurations
and average the solvation free energy results to eliminate the
configuration dependence. All simulations are performed at $T=\SI{298}{K}$
and $P=\SI{1}{atm}$. The equations of motion are integrated using the
stochastic dynamics equation. The Parrinello-Rahman
barostat\cite{RN638} is used to keep pressure constant. In the
calculation of solvation free energy, initial configurations in these
simulations are first equilibrated by performing an energy minimization
using the steepest descent algorithm, followed by a \SI{2}{ns} simulation in
canonical ensemble and \SI{5}{ns} simulation in isothermal-isobaric ensemble.
Then, for each solvent, \SI{40}{ns} production simulations are
performed.
\subsection {CG mapping and potentials}
\subsubsection {CG mapping}
\begin{figure}[h]
	\includegraphics[scale=0.2]{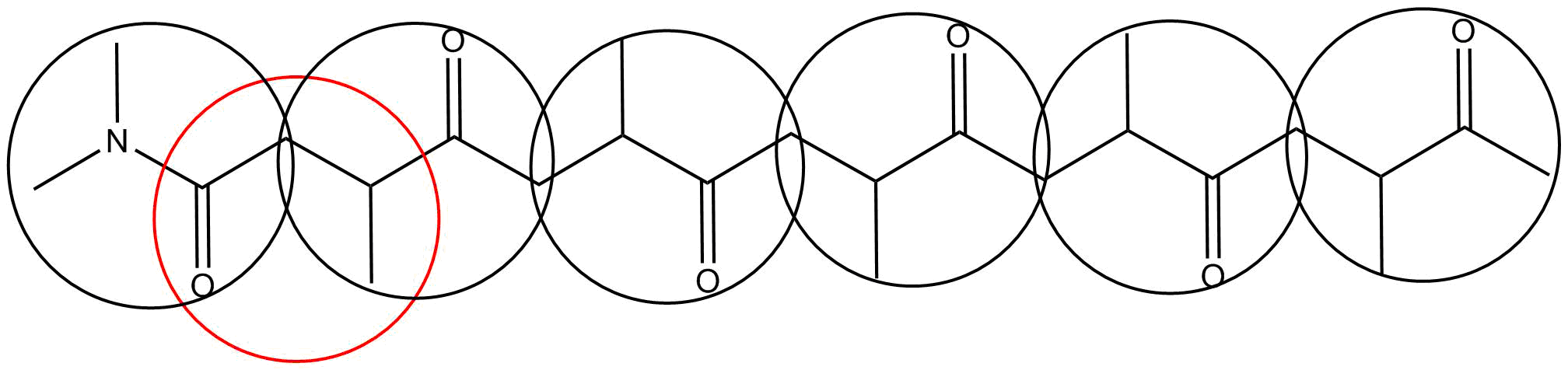}
	\includegraphics[scale=0.8]{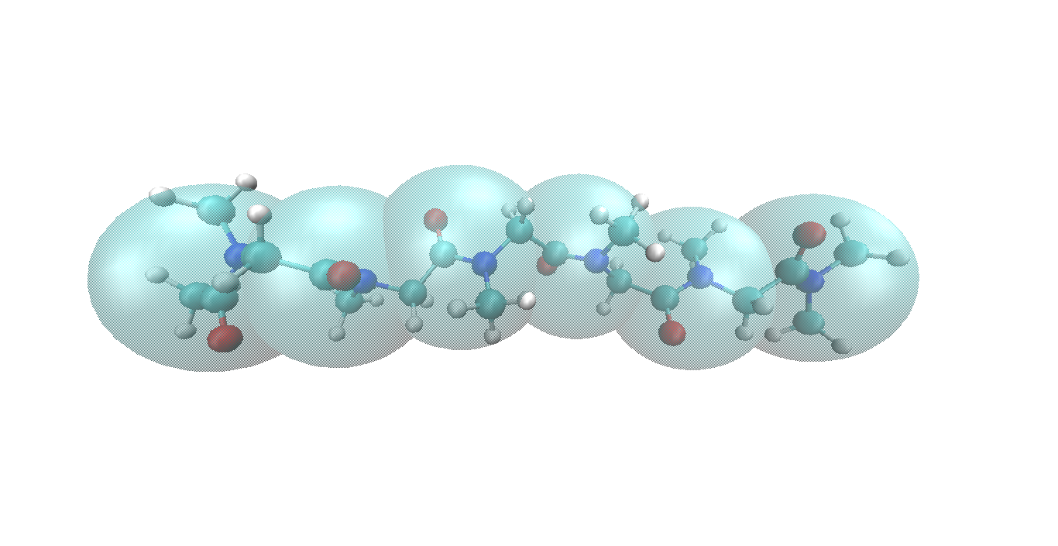}
	\caption{CG scheme (bottom) and its projection on the two-dimensional
		space (top) of poly ($\alpha$-peptoid) backbone (polysarcosine). The spheres
		(black circles in the two-dimensional projection) depict CG beads. The
		chemical structure of polysarcosine monomer is shown in the red circle.}
	\label{fig:scheme1}
\end{figure}

CG mapping from the atomic to coarse scale is not unique, and a 
mapping scheme can affect both predicting power and computation
efficiency of the CG model. In our CG model of the polysarcosine (poly
($\alpha$-peptoid) backbone), we define CG beads as shown in Figure  \ref{fig:scheme1}. 
Note that each CG
bead includes a half of the \ce{CH2} group and has mass of
71.076 (relative atomic mass), which is very close to that in the
original MARTINI FF (each CG bead has the mass of 72). This makes the
proposed CG mapping compatible with the MARTINI FF. Similar to the
polybutadiene CG model,\cite{RN478} we center beads at the
geometric center of the C-N bond. This choice of the bead placement
 allows the CG bond
potential to be fitted with a simple analytical form, as described below. The
solvents are modeled with CG ``solvent'' beads using the same degree of coarse-graining as in MARTINI FF.  
%

\begin{figure}[h]
	\includegraphics[scale=0.8]{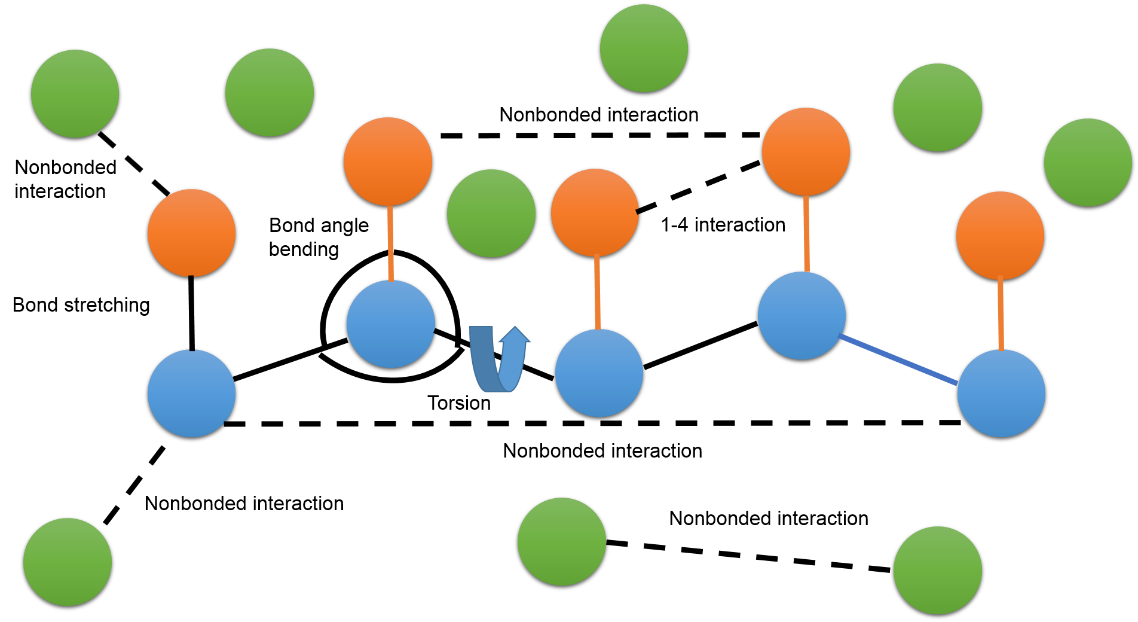}
	\caption{Interactions in  the CG model of poly ($\alpha$-peptoid) molecule
		with side chains in solvent. The CG backbone beads are labelled as blue
		circles, the CG side chain beads are orange, and the solvent beads are green.}
	\label{fig:scheme2}
\end{figure}

%
%\subsubsection{CG potentials and simulations}\label{CG_potentials}
%In the proposed CG model, we use the same 
%potential functions as in MARTINI FF. 
Figure \ref{fig:scheme2} schematically describes all
types of interactions between CG beads for a typical CG poly ($\alpha$-peptoid)
molecule with side chains in a solvent.
The total
potential energy \(U^{\text{CG}}\) due to pair and many-body
interactions between CG beads, representing the backbone and side chains,
consists of the bonded and nonbonded potentials:
\begin{equation}
U^{\text{CG}} = \sum U_{\text{bonded}}^{\text{CG}} + \sum U_{\text{nonbonded}}^{\text{CG}}.
\label{eq:eq1}
\end{equation}
In the next two sections, we discuss parameterization of these potentials.
\subsubsection {Parameterization of the bonded potential in CG FF}
The bonded potential, \(U_{\text{bonded}}^{\text{CG}}\), can be divided
into bond stretching, \(U_{\text{bond}}^{\text{CG}}\), bond angle
bending, \(U_{\text{angle}}^{\text{CG}}\), and torsion,
\(U_{\text{torsion}}^{\text{CG}}\), terms:
\begin{equation}
U_{\text{bonded}}^{\text{CG}}\left( r,\theta,\varphi \right) = U_{\text{bond}}^{\text{CG}}\left( r \right) + U_{\text{angle}}^{\text{CG}}\left( \theta \right) + U_{\text{torsion}}^{\text{CG}}\left( \varphi \right).
\label{eq:eq2}
\end{equation}
where, \(r\), \(\theta\), and \(\varphi\) are the bond length, bond
angle, and dihedral angle, respectively.
We compute \(U_{\text{bond}}^{\text{CG}}(r)\),
\(U_{\text{angle}}^{\text{CG}}(\theta)\), and
\(U_{\text{torsion}}^{\text{CG}}\left( \varphi \right)\) from atomic
simulations using the Boltzmann inversion method\cite{RN2} as:
\begin{equation}
U_{\text{bond}}^{\text{CG}}\left( r \right) = - k_{B} T \text{ln}\left( \frac{P_{\text{bond}}^{\text{CG}}\left( r \right)}{r^{2}} \right) + C_{r},
\end{equation}
\begin{equation}
U_{\text{angle}}^{\text{CG}}\left( \theta \right) = - k_{B} T \text{ln}\left( \frac{P_{\text{angle}}^{\text{CG}}\left( \theta \right)}{\text{sin$\theta$}} \right) + C_{\theta},
\end{equation}
\begin{equation}
U_{\text{torsion}}^{\text{CG}}\left( \varphi \right) = - k_{B} T \text{ln}\left( P_{\text{torsion}}^{\text{CG}}\left( \varphi \right) \right) + C_{\varphi}.
\end{equation}
Here, \(P_{\text{bond}}^{\text{CG}}\left( r \right)\),
\(P_{\text{angle}}^{\text{CG}}\left( \theta \right),\) and
\(P_{\text{torsion}}^{\text{CG}}\left( \varphi \right)\) are the
probability density functions (PDFs) of the bond lengths, bond angles,
and dihedral angles between CG beads. The positions of CG beads
are found from the atomic MD simulation of a given peptoid in a
desired solvent. We apply the Boltzmann inversion method to compute CG
bonded potentials for polysarcosine and poly (N-(2-carboxyethyl) glycine)
in acetonitrile.
The CG model of  
poly (N-pentyl glycine) is parameterized by
transferring bonded potential for interaction between backbone beads
from the polysarcosine CG model and using MARTINI FF for the bonded
potential between backbone and sidechain beads. The transferability of
bonded potentials 
\(P_{\text{bond}}^{\text{CG}}\left( r \right)\),
\(P_{\text{angle}}^{\text{CG}}\left( \theta \right),\) and
\(P_{\text{torsion}}^{\text{CG}}\left( \varphi \right)\) between
backbone beads is based on the assumption that these potentials are the same for the considered peptoids in any  good
solvent.

\begin{figure}
\includegraphics[scale=0.35]{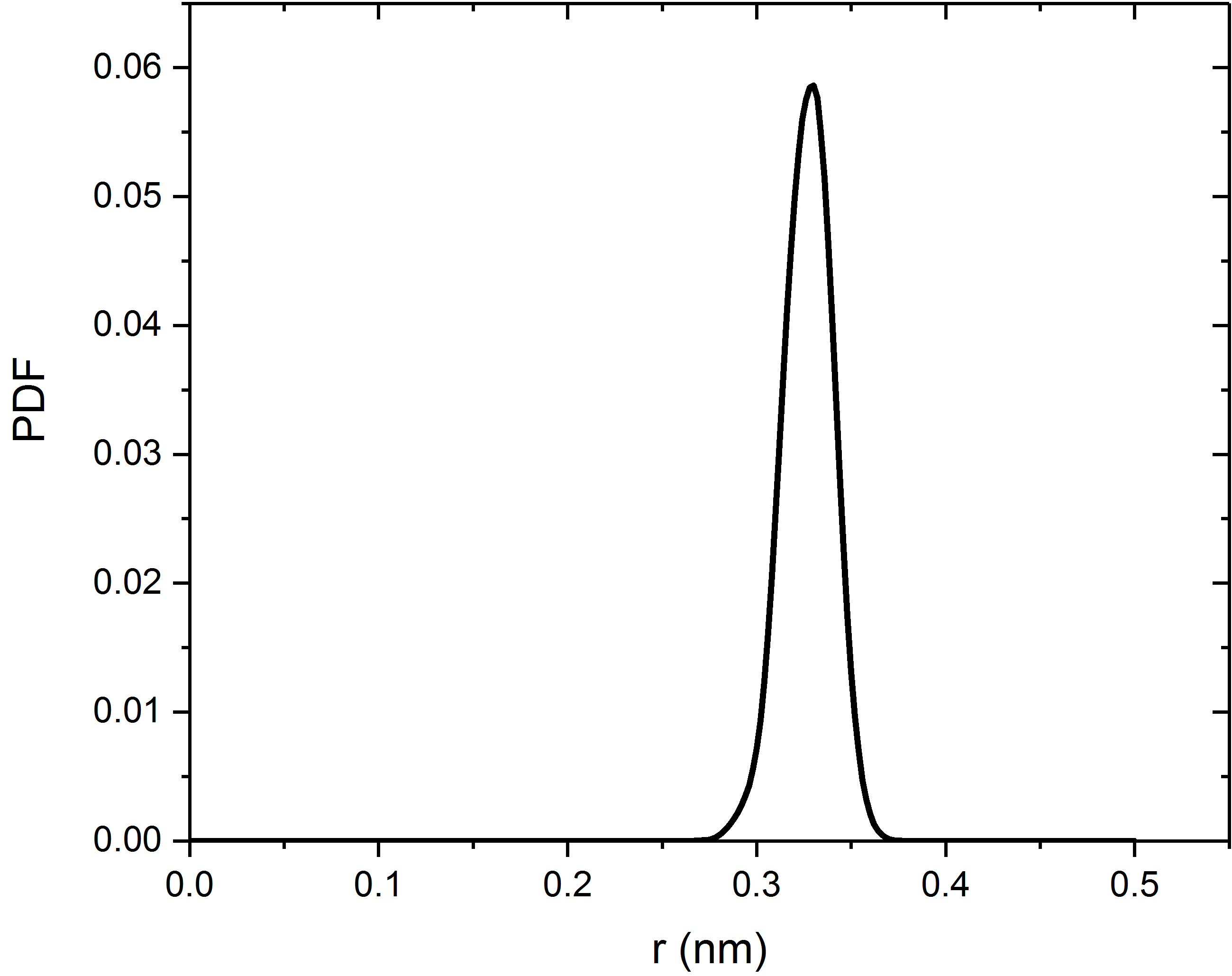}
\includegraphics[scale=0.35]{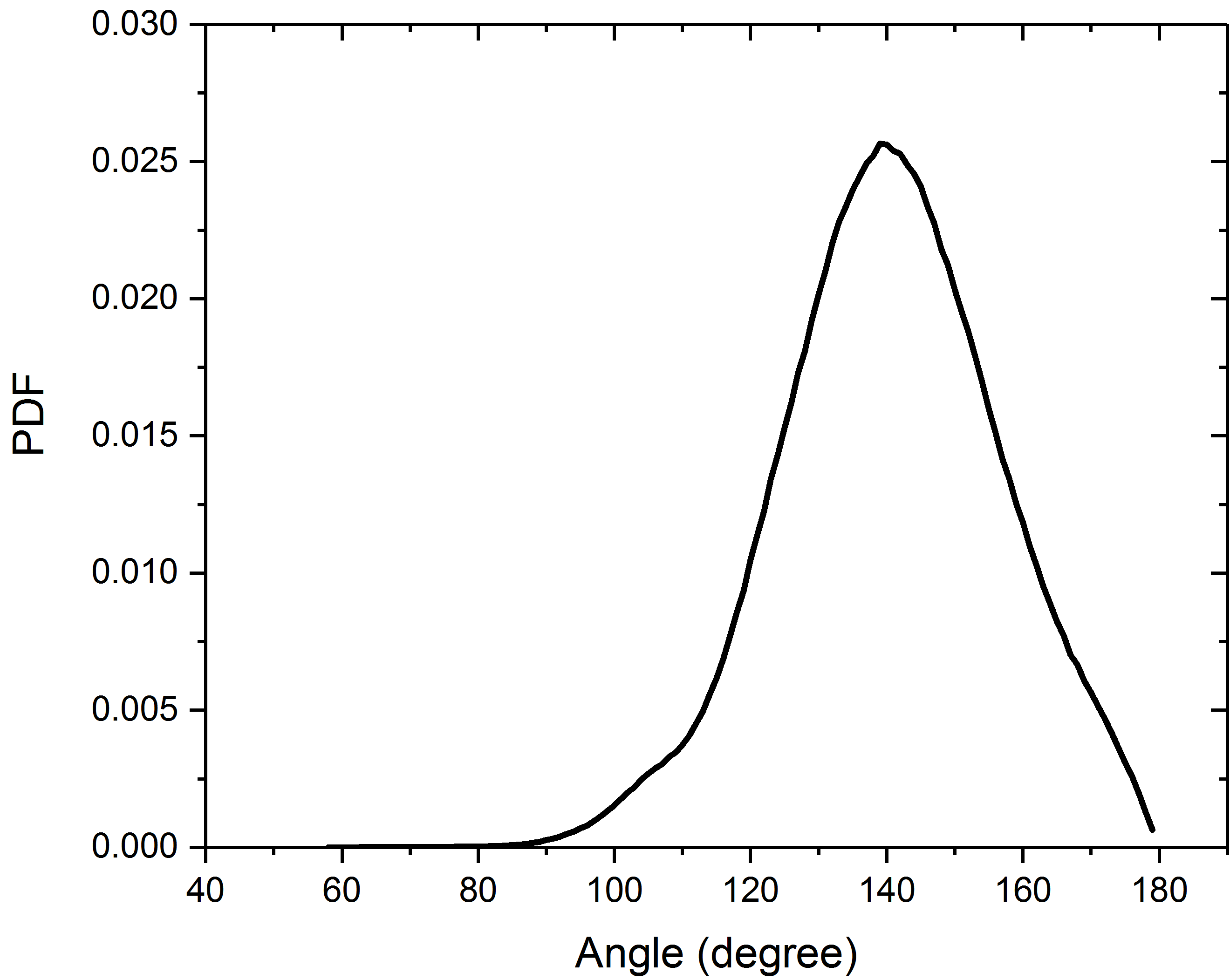}
\includegraphics[scale=0.35]{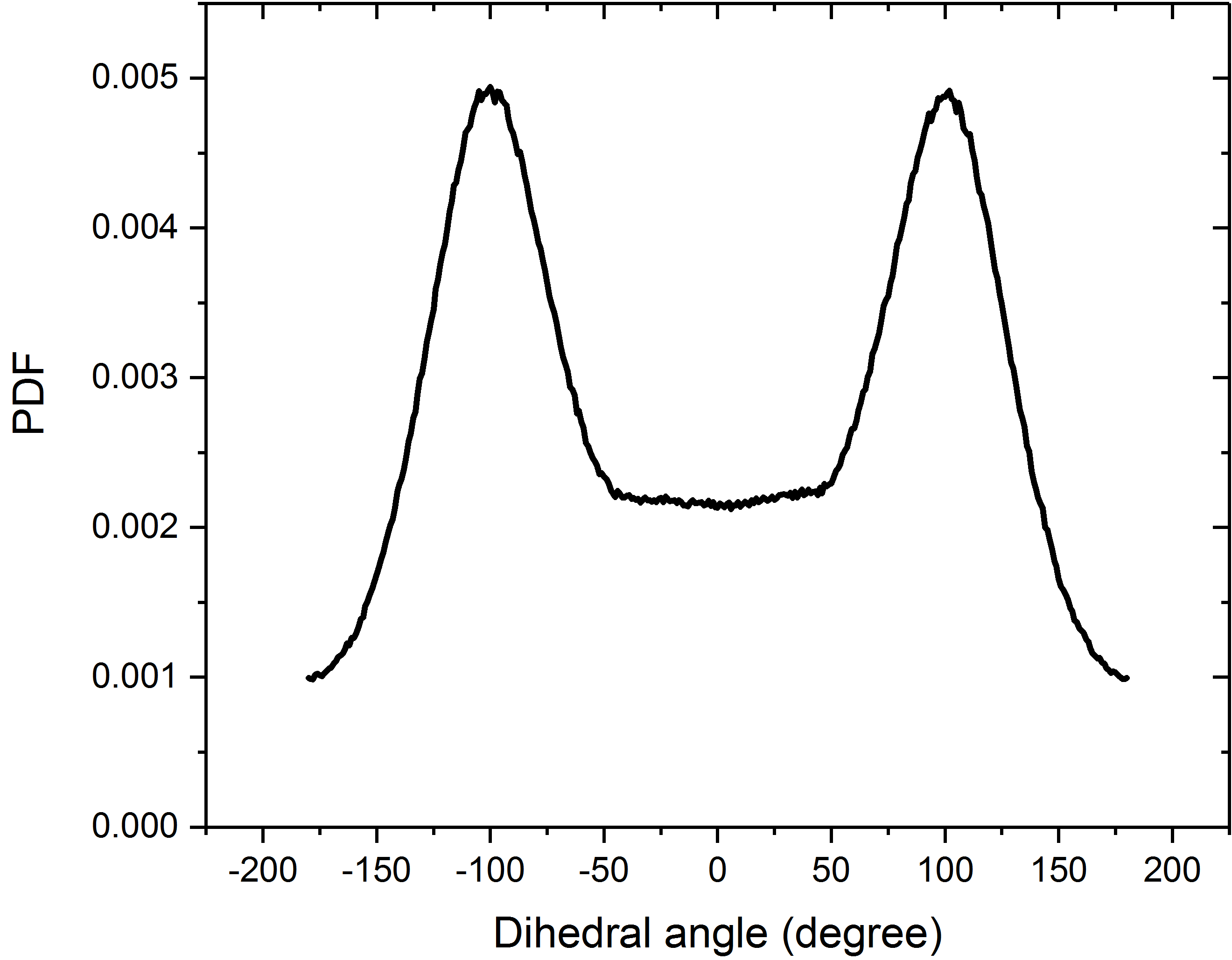}
\caption{PDFs of (a) bond length, (b) bond angle, and (c) dihedral angle of the CG polysarcosine in acetonitrile obtained from atomic simulations.} 
\label{fig:PDF}
\end{figure}

In Figure \ref{fig:PDF}a, we see that \(P_{\text{bond}}^{\text{CG}}\left( r \right)\) between
backbone beads of polysarcosine in acetonitrile has
Gaussian distribution. Therefore,
the bond potential between the backbone beads can be accurately
represented with the harmonic function:
\begin{equation}
U_{\text{bond}}^{\text{CG}}\left( r \right) = \frac{1}{2}k_{r}(r - r_{0})^{2}.
\end{equation}

\begin{figure}
\includegraphics[scale=0.9]{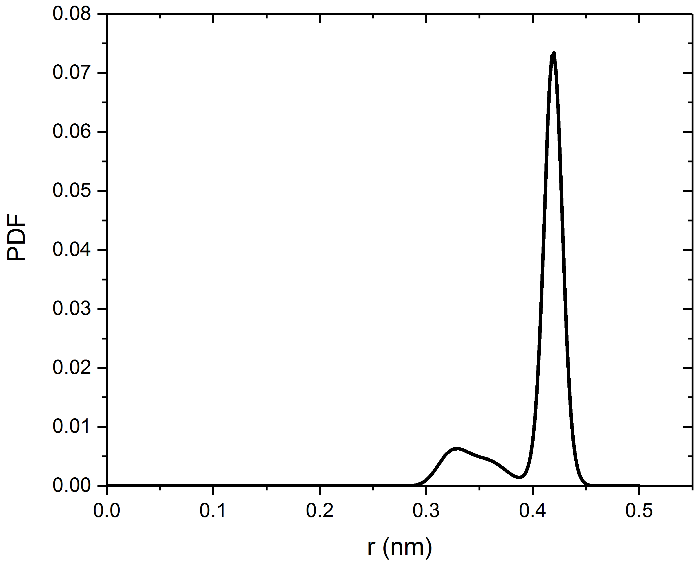}
\caption{The bond PDF between backbone and sidechain CG beads for poly (N-(2-carboxyethyl) glycine) in acetonitrile obtained by analyzing the trajectory in atomic simulations.}
\label{fig:PDF2}
\end{figure}

Figure \ref{fig:PDF2} demonstrates that for poly (N-(2-carboxyethyl) glycine) chain in acetonitrile, \(P_{\text{bond}}^{\text{CG}}\left( r \right)\) between backbone and
side chain has a bimodal distribution. Therefore,
\(U_{\text{bond}}^{\text{CG}}\left( r \right)\) for this peptoid does not have a simple
representation, and we prescribe it in a tabulated form.  We also find that PDFs \(P_{\text{angle}}^{\text{CG}}\left( \theta \right),\) in polysarcosine (Figure  \ref{fig:PDF}b) and other petoids (for bonds between two backbone and backbone and side-chain beads) have non-Gaussian distributions. Therefore, the
bond angle potentials \(U_{\text{angle}}^{\text{CG}}\) are given in the tabulated form.

We find that the torsion potential between the backbone beads has a symmetric bimodal distribution and, therefore, can be
well approximated with the analytical function \cite{RN2000}
\begin{equation}
U_{\text{torsion}}^{\text{CG}}\left( \varphi \right) = \sum_{n = 0}^{5}( - 1)^{n}k_{n}(cos\varphi)^{n}.
\end{equation}
All CG bonded
interactions are listed in Tables \ref{Tab:stretching}--\ref{tab:torsion}.
 Similar to Huang's CG poly(methyl methacrylate) model\cite{RN688}, we disregard the
torsion potentials between backbone and side-chain beads and two neighbor side-chain 
beads. To avoid the overlap between beads due to the disregarded
torsion potential, the ``1-4" nonbonded interactions between backbone and
side-chain beads and two side-chain beads (see Figure \ref{fig:scheme2}) are included in the form of  
 the LJ 12-6 potential
  \begin{equation}
 \label{LJpotential}
U_{\text{nonbonded}}^{\text{CG}} = 4\varepsilon\left[\left( \frac{\sigma}{r} \right)^{12} - \left( \frac{\sigma}{r} \right)^{6}\right].
\end{equation}

\begin{figure}
\includegraphics[scale=0.35]{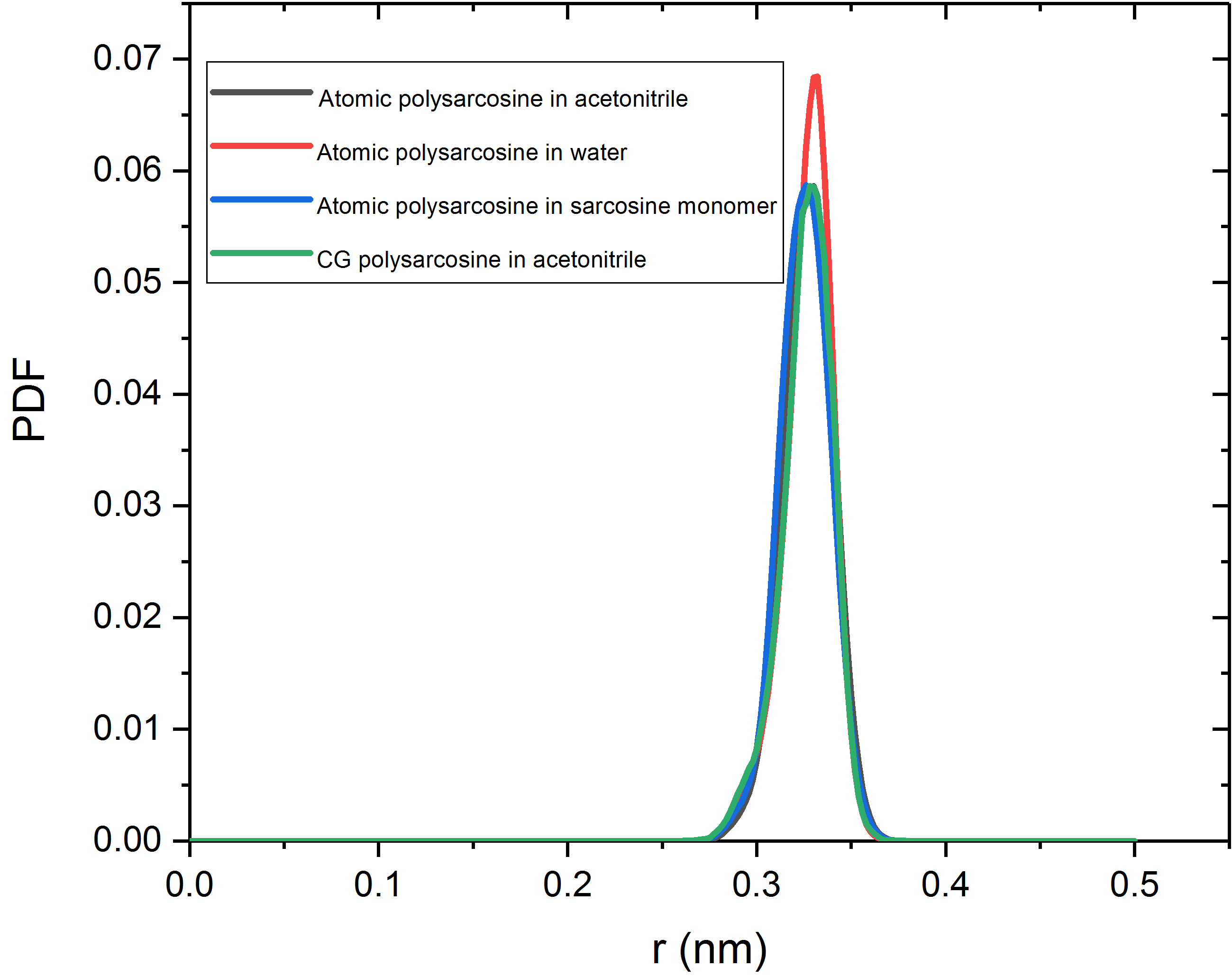}
\includegraphics[scale=0.35]{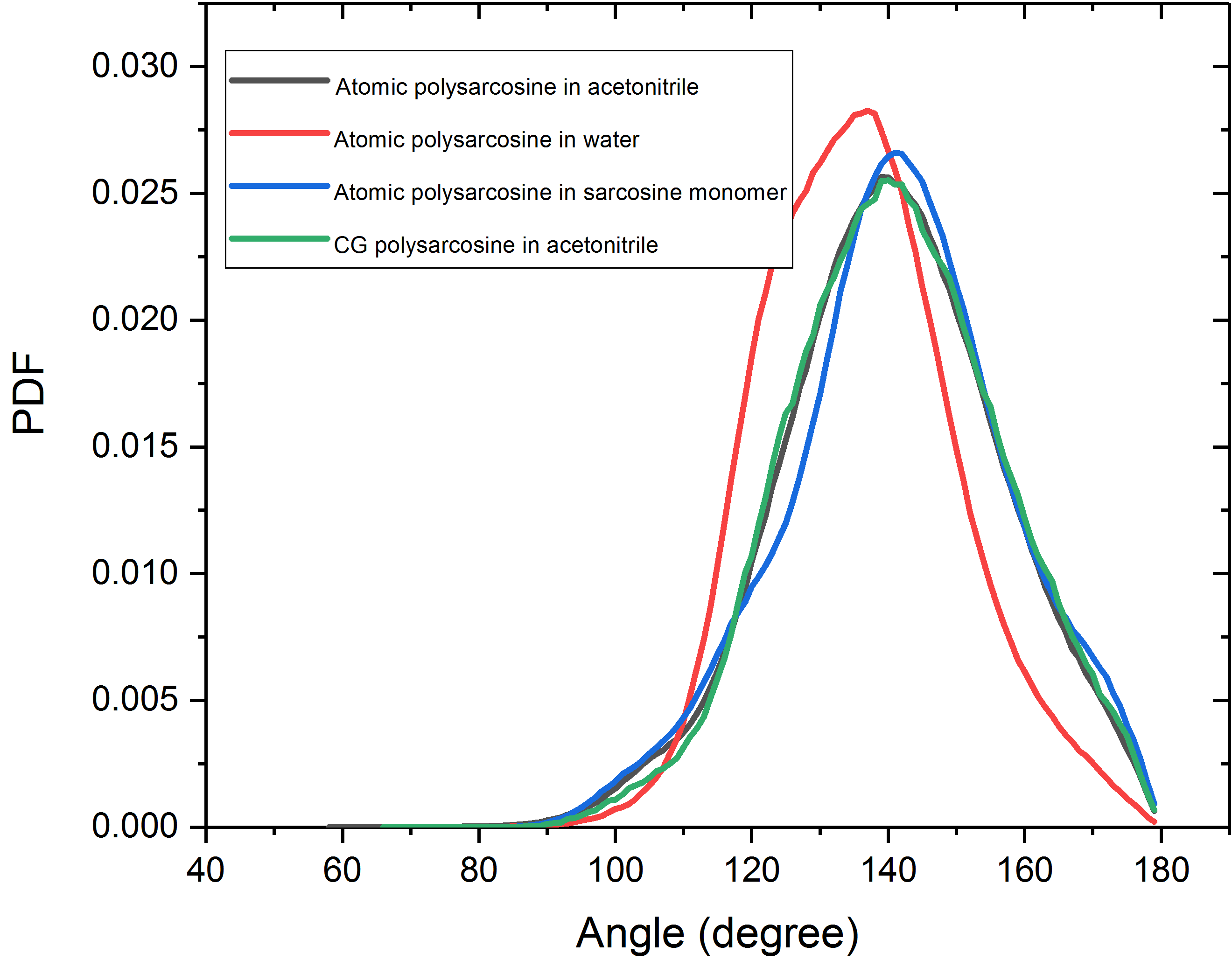}
\includegraphics[scale=0.35]{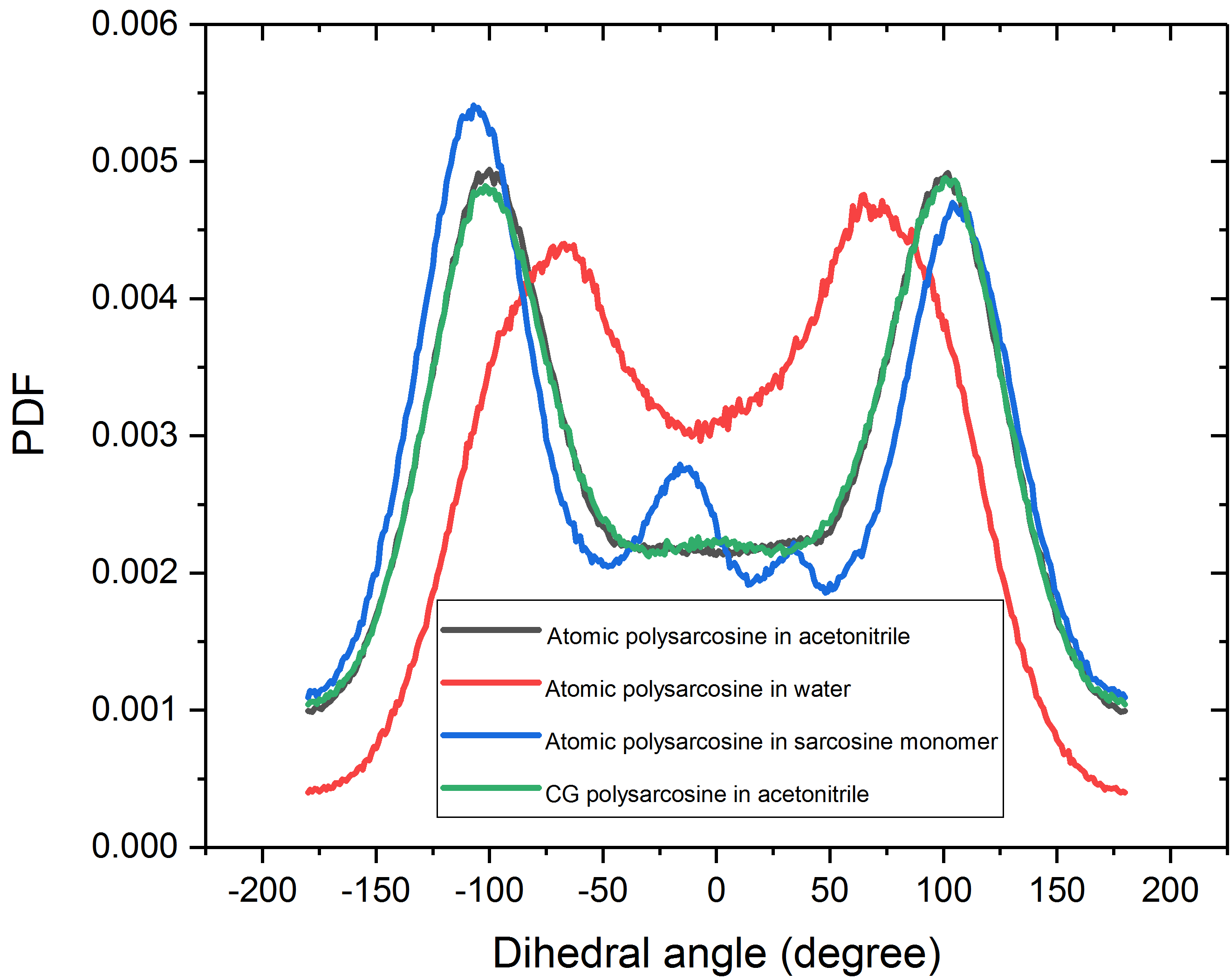}
\caption{(a) PDF of bond, (b)PDF of angle, (c)PDF of dihedral angle of CG polysarcosine in CG acetonitrile, water and sarcosine monomer obtained from atomic simulations. }
\label{fig:s3}
\end{figure}

To evaluate the dependence of local structures on solvent types, we
analyzed the atomic simulation trajectory of polysarcosine chain in a CG manner
(i.e., we determine locations of CG beads from atomic simulations as described in Figure \ref{fig:scheme1} and obtain the PDFs for the bond
length, bond angle, and dihedral angle of CG polysarcosine chain in
acetonitrile, water and sarcosine monomer. Figure \ref{fig:s3}, shows that the proposed CG model can reproduce the local conformation of polysarcosine in acetonitrile. This figure also demonstrates that PDFs of bond length and angle are practically  independent of the solvent type. On the other hand, the PDF of dihedral angle
show strong dependence on the solvent type. This implies that the CG
bond and angle potentials are transferable in various solvent, but
the torsion potential may not be transferable. 

% According to Nathan's comment, we need all the CG results for the three solvents, Currently I only have one.
\begin{table}
\caption{CG Bond stretching potential for the polysarcosine between sarcosine (backbone) CG beads (denoted as PA) and  acetic group
(side chain) beads (denoted as P3). This naming convention follows the MARTINI FF. }
\begin{tabular}{ccc}
\hline
Bond stretching & \(k_{\text{r}}\)(KJ/mol) & \(r_{\text{0}}\)(nm) \\
\hline
PA-PA & 17000 & 0.332 \\
PA-P3 & Tabulated (see Figure S1) \\
\hline
\label{Tab:stretching}
\end{tabular}
\end{table}
\begin{table}
\caption{CG Bond angle bending potential for polysarcosine.}
\begin{tabular}{ccc}
\hline
 Bond angle bending & k(KJ/mol) & $\theta$(degree) \\
\hline
PA-PA-PA & Tabulated (see Figure S1) \\
PA-PA-P3 & Tabulated (see Figure S1) \\
\hline
\end{tabular}
\end{table}

\begin{table}
\caption{CG torsion potential for polysarcosine.}
\begin{tabular}{ccccccc}
\hline
Torsion & \(k_{0}\)(KJ/mol)
& \(k_{1}\)(KJ/mol) &
\(k_{2}\)(KJ/mol) & \(k_{3}\)(KJ/mol) &
\(k_{4}\)(KJ/mol) & \(k_{5}\)(KJ/mol) \\
\hline
PA-PA-PA-PA & 0.19054 & $-1.59509$ & 3.61401 & 3.56366 & $-0.76257$ &
$-0.62289$ \\
\hline
\end{tabular}
\label{tab:torsion}
\end{table}
\subsection {Parametrization of the nonbonded potentials in the CG FF}
Following the MARTINI FF, for the nonbonded interaction potential we
use the LJ 12-6 potential (\ref{LJpotential}).
In the LJ 12-6 potential, the parameter \(\sigma\)
represents the closest distance between two CG beads and
\(\varepsilon\) is the strength of their interaction.
The nonbonded interactions act between beads of a peptoid
chain and solvent, between beads of different chains, and also between
beads on the same chain separated by three beads or more.

%For a typical poly $\alpha$-peptoid molecule with side chains in a solvent, the CG nonbonded potential can be formulated as
%\begin{equation}
%U_{\text{nonbonded}}^{\text{CG}}\left( \varepsilon,\sigma \right) = U_{\text{bb}}^{\text{CG}}\left( \varepsilon,\sigma \right) + U_{\text{bs}}^{\text{CG}}\left( \varepsilon,\sigma \right) + U_{bo}^{\text{CG}}\left( \varepsilon,\sigma \right) + U_{\text{ss}}^{\text{CG}}\left( \varepsilon,\sigma \right) + U_{\text{so}}^{\text{CG}}\left( \varepsilon,\sigma \right) + U_{\text{oo}}^{\text{CG}}\left( \varepsilon,\sigma \right).
%\end{equation}
%Here, the subscript ``b'' denotes a backbone bead, ``s''  stands for a sidechain bead, and ``o'' represents a solvent bead.

In the original MARTINI FF, the magnitude of \(\sigma\) depends on the
degree of coarse graining and is set to \(\sigma =\) \SI{0.47}{nm} for the 4:1
CG mapping (four heavy atoms mapped to one ``large'' CG bead) and \(\sigma =\) \SI{0.43}{nm} 
for the 3:1 CG mapping (``small'' beads). The \(\varepsilon\) value depends on the
types of CG beads. The interactions between CG beads are divided into
four main types: polar (P), nonpolar (N), apolar (C), and charged (Q).
Each main type has five subtypes, which are distinguished by the
hydrogen-bonding capabilities or the degree of polarity. The range of \(\varepsilon\) is from 2.0 to \SI{5.6}{KJ/mol}.\cite{RN274}
For convenience, it is divided in 10 levels in the MARTINI
FF. Each level is ``fine-tuned'' to reproduce the experimentally
observed solubilities. 
%The standard MARTINI FF gives \(\sigma\) and\(\varepsilon\) for some solvents. 
In this work, we
consider four solvents: water, hexane, 1-octanol, and
acetonitrile. Depending on types of
interacting beads, we compute \(\sigma\) and \(\varepsilon\) by matching
Rg or solvation free energy. 
 For some solvents, \(\sigma\) and \(\varepsilon\)
can be found in the original MARTINI FF. 
All nonbonded interaction pairs in this study, as well as the methods used for estimating
the \(\sigma\) and \(\varepsilon\) parameters for each pair, are listed in Table \ref{tab:IntPairs}.

\begin{table}
\caption{CG nonbonded interaction pairs in the poly ($\alpha$-peptoid) solution. Sarcosine=PA, water=P4, Acetonitrile=Snd1, \emph{n}-butyl group=P1, \emph{n}-butanol group=C1, Hexane=SC1, Acetic group=P3. R=obtained by matching $R_g$, S=obtained by matching solvation
free energy, M=original MARTINI FF. TBD=not discussed in this work. For
1-octonal CG model, it is a combination of CG \emph{n}-butyl group and
CG \emph{n}-butanol group.} 
\begin{tabular}{cccccccc}
\hline
Pair & PA & P4 & SNd1 & C1 & P1 & SC1 & P3 \\
\hline
PA & R & S & S & S & S & S & S \\
P4 & & M & S & M & M & M & M \\
SNd1 & & & S & TBD & TBD & TBD & S \\
C1 & & & & M & M & M & M \\
P1 & & & & & M & M & M \\
SC1 & & & & & & M & M \\
P3 & & & & & & & M \\
\hline
\end{tabular}
\label{tab:IntPairs}
\end{table}

\subsection{Parametrization of the nonbonded potentials of CG acetonitrile
bead}

\begin{figure}
\includegraphics[scale=1]{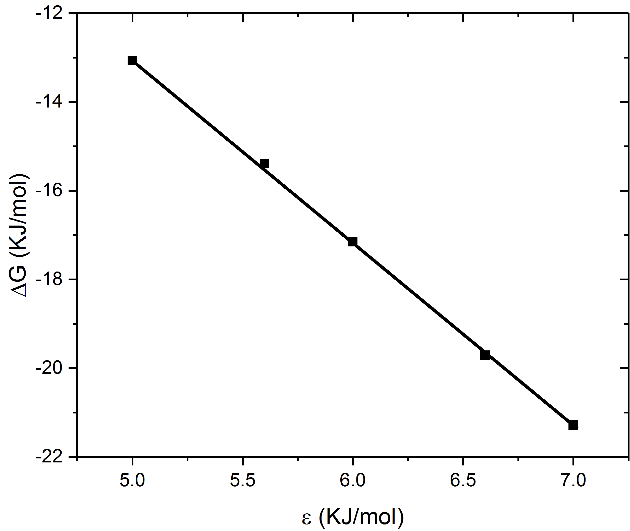}
\caption{The self-solvation free energy of CG acetonitrile versus
interaction parameter \(\varepsilon_{SNd1 - SNd1}\).}
\label{figure3}
\end{figure}

As shown in Table \ref{tab:IntPairs}, for all considered solvents, except acetonitrile,
\(\sigma\) and \(\varepsilon\) are given in the original MARTINI CG FF
and its extensions. Note that MARTINI CG FF uses the P4 type bead for
water, SC1 for hexane, and P1+C1 for 1-octanol. 
Because of the acetonitrile chemical properties, the type of 
acetonitrile beads should be one of N or P subtypes. We find that none of these subtype beads can reproduce the solvation free energy computed from atomic simulation. 
%Since an acetonitrile molecule includes only three heavy atoms, following the MARTINI FF rule, \cite{RN274} we treat it as a ``small'' CG bead with \(\sigma =\) 0.43 nm and \(\varepsilon = 0.75\ \varepsilon_{0}\), where \(\varepsilon_{0}\) is the common nonbonded interaction. 
Therefore, we define
a new CG bead subtype SNd1 for CG acetonitrile bead with $\sigma_{SNd1 - SNd1} = \SI{0.43}{nm}$
(according to the MARTINI rule for interactions between beads made of three heavy atoms). The parameter \(\varepsilon_{SNd1 - SNd1}\)  for the potential between acetonitrile beads is obtained by computing solvation free energy as a function of $\varepsilon$. Figure \ref{figure3} presents the relationship
between $\varepsilon$ and the corresponding solvation
free energy for CG acetonitrile solvated in acetonitrile solvent. The
solvation free energy linearly decreases with increasing
$\varepsilon_{SNd1 - SNd1}$. We obtain
$\varepsilon_{SNd1 - SNd1} = \SI{6.570}{KJ/mol}$ to reproduce the 
solvation free energy
$\mathrm{\Delta}G = -19.51 \pm \SI{0.04}{KJ/mol}$ calculated from the atomic simulation. Note that this value exceeds the range of 2--$\SI{5.6}{KJ/mol}$
for \(\varepsilon\) in the original MARTINI FF. 

Next, we parameterize the LJ potential for interaction between SNd1 (acetonitrile) and P4 (water) beads. The $\sigma_{SNd1 - P4}$ value for this potential is set to $\SI{0.47}{nm}$.  The
interaction parameter $\varepsilon_{SNd1 - P4} = \SI{4.520}{KJ/mol}$ is
found as above to match the solvation free energy of water in acetonitrile,
$\Delta G = -13.75 \pm \SI{0.12}{KJ/mol}$, found from the atomic simulation. Also, our atomic simulation results are in good
agreement with the experimental values \(\mathrm{\Delta}G = -20.29 \pm 0.84 \)
KJ/mol for acetonitrile self-solvation free energy and \(\mathrm{\Delta}G = -16.23 \pm 2.51 \)
KJ/mol for water in acetonitrile \cite{RN508}.

\subsection{Parametrization of the nonbonded potentials between CG
polysarcosine and solvent beads.}

\begin{figure}
\includegraphics[scale=1]{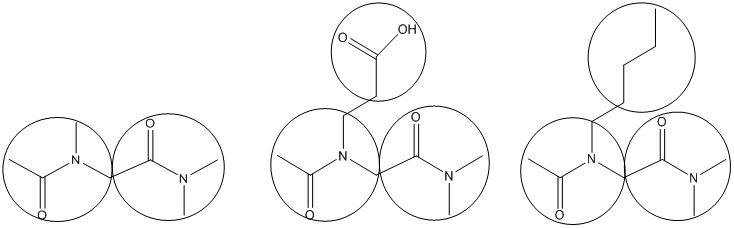}
\caption{The atomic and CG chemical structure of polysarcosine, Poly
(N-(2-carboxyethyl) glycine), and Poly (N-pentyl glycine) monomers with
terminal groups.}
\label{fig:polysarcisine}
\end{figure}

 We determine the CG potentials between considered peptoids and any
solvent in this section.
Specifically, we compute the CG nonbonded potentials between
polysarcosine monomer and water, 1-octanol, acetonitrile or hexane. In
the original MARTINI CG FF, the glycine residue, which is similar to
sarcosine, is labelled as type P5\cite{RN22}. Using the MARTINI FF for interaction between P5 bead and water, we obtain the hydration free energy $\SI{-40}{kJ/mol}$, which is significantly larger than  
the hydration free energy of
polysarcosine monomer ($\SI{-48.16}{kJ/mol}$) computed from the atomic simulation. Therefore, we define a new bead type,
PA, for the CG polysarcosine monomer bead and compute the nonbonded
interaction parameters between PA and solvent beads to reproduce the
hydration free energy of polysarcosine monomer (see Figure \ref{fig:polysarcisine} for
chemical structure details) in the atomic simulation.
\begin{figure}
\includegraphics[scale=1]{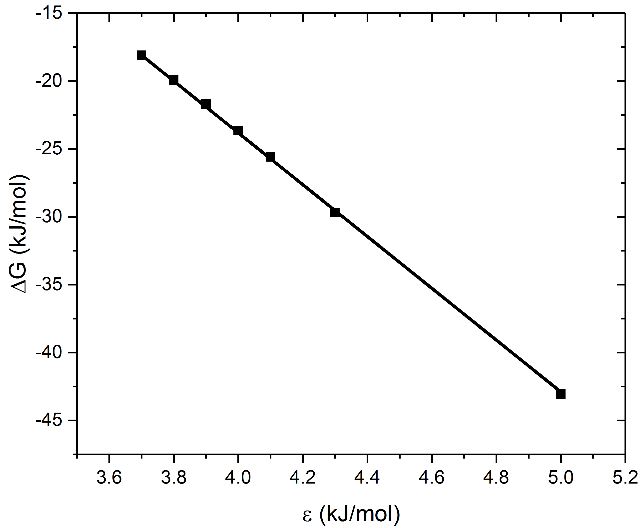}
\caption{The hydration free energy of CG polysarcosine monomer versus
\(\varepsilon\)}
\label{fig:figure5}
\end{figure}
Figure \ref{fig:figure5} shows that the hydration free energy linearly decreases as
\(\varepsilon\) increases. We find \(\varepsilon_{PA - P4}\) = $\SI{5.264}{kJ/mol}$ that
reproduces the desired hydration free energy by linear fitting. This \(\varepsilon\)
value is smaller than the interaction~($\SI{5.6}{kJ/mol})$ between glycine 
and water beads in the MARTINI CG FF. Also, it is consistent with the fact
that the polarity of a sarcosine molecule is smaller than the polarity
of a glycine molecule. We also use the same method to obtain
\(\varepsilon\) in the potentials acting between polysarcosine 
and other solvents CG beads, including 1-octanol, acetonitrile and hexane. Figure
\ref{fig:figure6} shows the solvation free energy of polysarcosine monomer in hexane as
a function of \(\varepsilon\). From this figure, we find
\(\varepsilon_{PA - SC1} = \SI{3.197}{KJ/mol}\). For 1-octanol, the estimation
of \(\varepsilon\) is complicated because in MARTINI FF, 1-octanol molecules comprise
of two CG beads with different types (P1 and C1). We
first determine \(\varepsilon_{PA - C1}\) in the potential acting
between PA and C1 beads. 
Next, we find $\varepsilon_{PA - P1} = \SI{4.851}{kJ/mol}$  by matching solvation free energy in the corresponding atomic simulation (The red one in Figure 9).  The solvation free energy of
polysarcosine monomer in four solvents obtained in CG and atomic
simulations are listed in Table \ref{tab:solvation}.

\begin{figure}
\includegraphics[width=3.5913in,height=2.92944in]{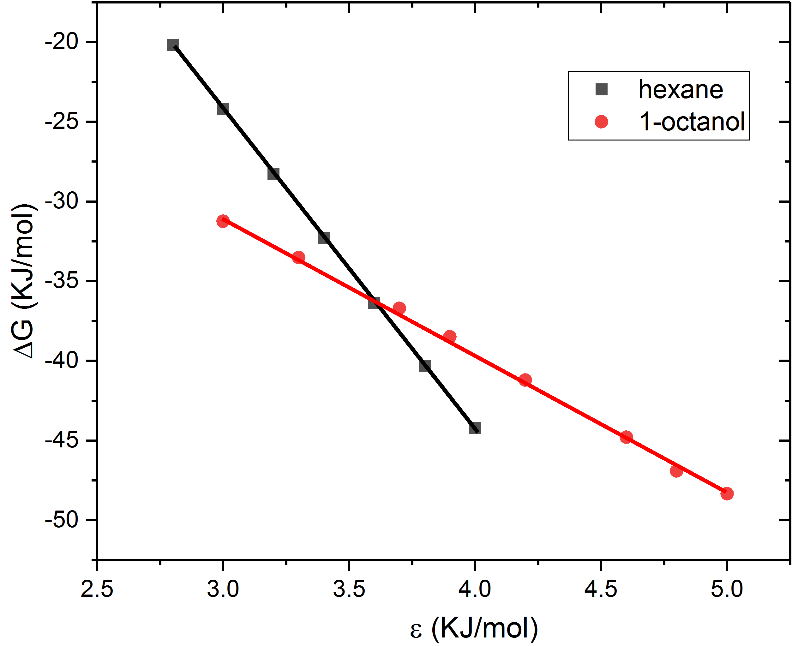}
\caption{The solvation free energy of polysarcosine monomer in
1-octanol and hexane vs interaction parameter \(\varepsilon\).}
\label{fig:figure6}
\end{figure}

\begin{table}
\caption{Solvation free energy of
polysarcosine monomer in various solvents computed from atomic and CG simulations.}
\begin{tabular}{ccccc}
\hline
$\Delta$G(KJ/mol) & $\Delta$\(G_{\text{acetonitrile}}\) & $\Delta$\(G_{\text{hydration}}\) & $\Delta$\(G_{\text{1-octanol}}\) & $\Delta$\(G_{\text{hexane}}\) \\
\hline
atomic & $-49.41\pm0.20$ & $-48.16\pm0.06$ & $-47.32\pm0.47$ &
$-28.21\pm0.08$ \\
CG & $-49.31\pm0.24$ & $-48.39\pm0.10$ & $-47.33\pm0.17$ &
$-28.27\pm0.14$ \\
\hline
\end{tabular}
\label{tab:solvation}
\end{table}

\subsection{Parametrization of the nonbonded potentials between CG peptoid
beads.}

\begin{figure}
\includegraphics[width=3.03333in,height=2.50315in]{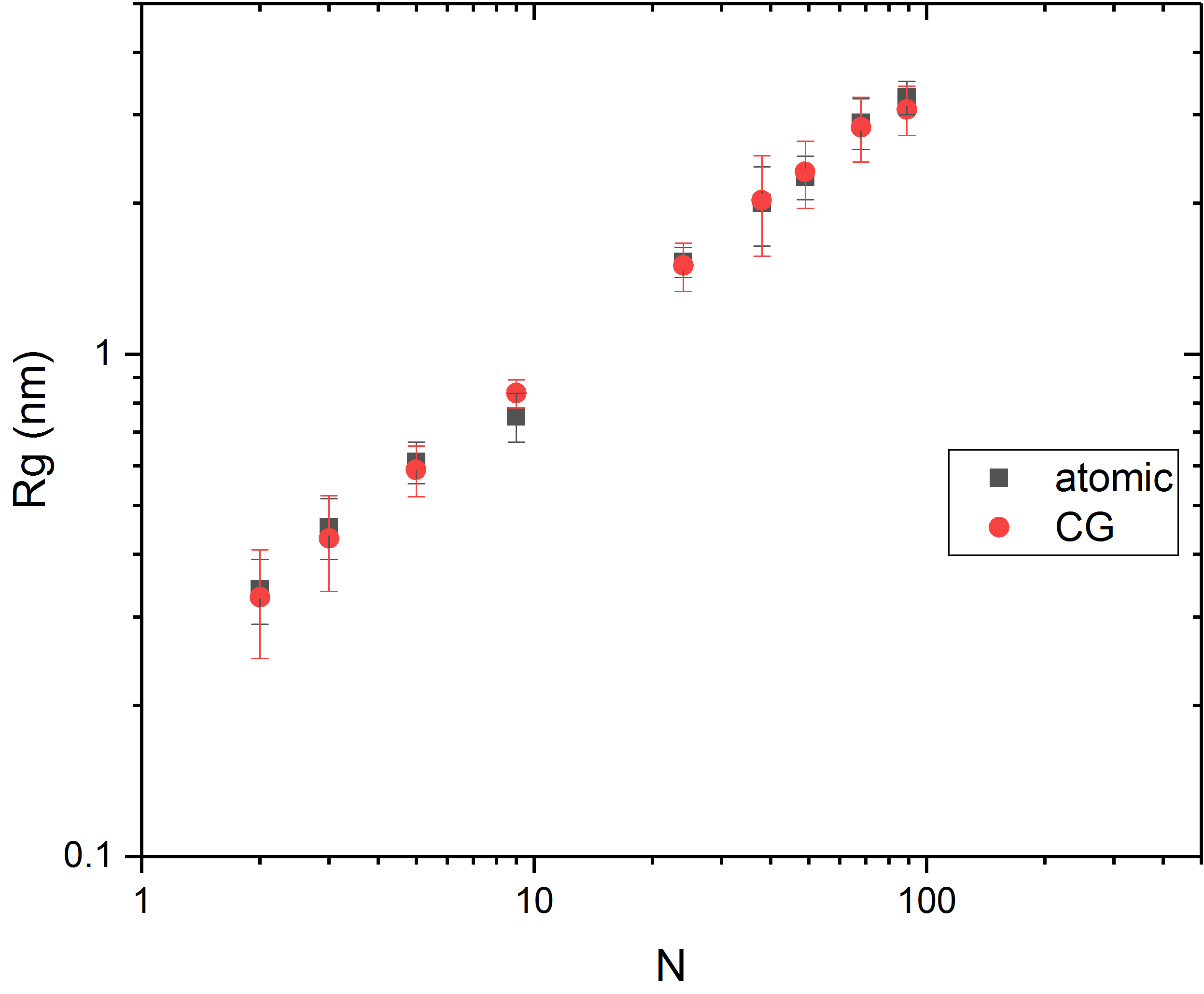}
\caption{$R_g$ of polysarcosine in acetonitrile with repeat units number N in
atomic and CG simulations.}
\label{fig:Rg}
\end{figure}

In the above section, we show how to parameterize the nonbonded
potential between poly ($\alpha$-peptoid) backbone (polysarcosine) and solvent
beads. Since the
intramolecular nonbonded interactions are excluded for polysarcosine chain
with length less than four repeat units, the interaction parameter between CG sarcosine repeat
units does not affect the solvation
free energy of polysarcosine monomer.  Here, we
compute \(\varepsilon_{PA - PA}\) for polysarcosine CG beads by matching
the $R_g$ for polysarcosine chain with 25 repeat units in acetonitrile obtained from the
atomic simulations.
% Although water is a very common solvent, existing
%experimental results\cite{RN207,RN664} and atomic 
%simulations\cite{RN515,RN525} are not conclusive on whether water
%is a good solvent for polysarcosine. 
We select acetonitrile in the calculation of $R_g$ because it was experimentally found
to be a good solvent for polysarcosine.\cite{RN207} 
Our atomic simulations show that $R_g$ has 
a power law behavior as a function of the number of repeat units (see Figure
\ref{fig:Rg}) with the scaling exponent 0.575, which, according to
Flory's theory, \cite{RN672} also confirms that acetonitrile is a good
solvent for polysarcosine. Then, we obtain \(\varepsilon_{PA - PA}\) = $\SI{5.6}{KJ/mol}$ to match
the $R_g$ of polysarcosine with 25 repeat units. The comparison of $R_g$ as a function of the repeat unit number, obtained from CG and atomic simulations, is shown in  Figure \ref{fig:Rg}. The good agreement demonstrates transferability of the nonbonded potential, i.e., \(\varepsilon_{PA - PA}\) obtained from a simulation of a chain with 25 repeat units accurately predicts $R_g$ of chains with other number of repeat units.

Theoretically, the solvation free energy of a polypeptoid with side chain in the CG model depends on the nonbonded potential between CG polypeptoid backbone and side chain. For example, the solvation free energy of Poly
(N-(2-carboxyethyl) glycine) is a function of  \(\varepsilon_{PA - P3}\). Therefore, the interaction parameter $\varepsilon$ could be
determined by matching the free energy in an atomic simulation of peptoid
with two repeat units (as was done to determine parameters in Figures \ref{fig:figure5}
and \ref{fig:figure6}). However, we performed CG simulations of poly
(N-(2-carboxyethyl) glycine) with two repeat units with several
\(\varepsilon_{PA - P3}\) values and found no obvious difference in the
resulting solvation free energy (see Table S1). Therefore, we set it to
\(\varepsilon_{PA - P3}=\SI{4}{KJ/mol}\), which corresponds to the level
III value in MARTINI FF. This parameter can be further optimized with long
poly ($\alpha$-peptoid) chain if needed. Parameters of all CG nonbonded
potentials are listed in Table S2. 

To further test transferability of the CG FF, we calculate the hydration free energy for
the polysarcosine, poly (N-(2-carboxyethyl) glycine) and poly N-pentyl
glycine and $R_g$ for polysarcosine and poly
(N-(2-carboxyethyl) glycine) in acetonitrile. The production simulation time is \SI{1000}{ns}
for calculating $R_g$ and \SI{50}{ns} for estimating the hydration free
energy. Additionally, we simulate a CG sequenced diblock peptoid chain with
100 beads in a binary mixture of water and
acetonitrile for \SI{1000}{ns} and calculate its $R_g$. In these simulations, the repeat unit for the diblock
polymer includes four sarcosine and one (N-(2-carboxyethyl) glycine) CG
beads and the acetonitrile concentration 
 varies from 0 to \SI{200}{mol/L}. In CG production simulations, constant
temperature and pressure are maintained using Nose-Hoover thermostat\cite{RN637} and Parrinello-Rahman barostat.\cite{RN638}  The LJ potential has the cutoff distance of
\SI{1.2}{nm} with smoothing after \SI{0.9}{nm}.\cite{RN639} Electrostatic
interactions are not included in the CG model. The time step in CG
simulations is \SI{10}{fs}. All the CG simulations are performed by GROMACS~
5.1.2. The results of these simulations are discussed in the following section.

\section {CG model transferability and simulations of peptoid folding}

\subsection {Transferability of the CG backbone parameters to other polypeptoids with respect to hydration free
energy}

Here, we study the transferability of the CG parameter
obtained for polysarcosine to other polypeptoids. We select poly
(N-(2-carboxyethyl) glycine) and poly (N-pentyl glycine) as
typical examples of peptoids with hydrophilic or hydrophobic side
chains and simulate them with atomic and CG models. The chemical structures of these peptoids are shown in Figure \ref{fig:polysarcisine}. Note that in the CG models, the nonbonded
interaction parameters for the new side chains are directly taken from the MARTINI FF. Table \ref{tab:energy_peptoids}
presents the calculated hydration free energies of poly
(N-(2-carboxyethyl) glycine) and poly (N-pentyl glycine) monomer in both
atomic and CG simulations. The difference of less than
5\% indicates 
good transferability of the CG backbone parameters obtained to other peptoids and compatibility of the proposed CG peptoid model with the MARTINI FF. 

\begin{table}
\caption{Hydration free energy of poly (N-(2-carboxyethyl) glycine)
monomer and poly (N-pentyl glycine) monomer in atomic and CG simulations}
\begin{tabular}{ccc}
\hline
$\Delta$\(G_{\text{hyd}}\)(KJ/mol) & Poly (N-(2-carboxyethyl) glycine) 
& Poly (N-pentyl glycine) \\
\hline
atomic & $-64.04\pm0.09$ & $-35.45\pm0.22$ \\
CG & $-66.30\pm0.17$ & $-38.85\pm0.18$ \\
\hline
\end{tabular}
\label{tab:energy_peptoids}
\end{table}
%\subsection {3.2 Transferability with respect to chain length}
%
\subsection {Transferability with respect to $R_g$}
According to the Flory theory, $R_g$ $\propto N^{v}$ in polymer solutions, where
\(N\) is the number of repeat units and \(\nu\) is the Flory
parameter.\cite{RN672} The parameter \(\nu\) is 0.59 for a
good solvent and 0.30 for a poor solvent. Figure \ref{fig:Rg} shows results of
atomic and CG simulations of polysarcosine in acetonitrile. We see
that $R_g$ in both the atomic and CG simulations increases with \(N\).
The $R_g$ obtained in CG simulations are in good agreement with the atomic
results. The fitted \(\nu\) from CG simulations is 0.591, which is
consistent with the Flory theory.\cite{RN672}

\begin{figure}
\includegraphics[width=3.06739in,height=2.53125in]{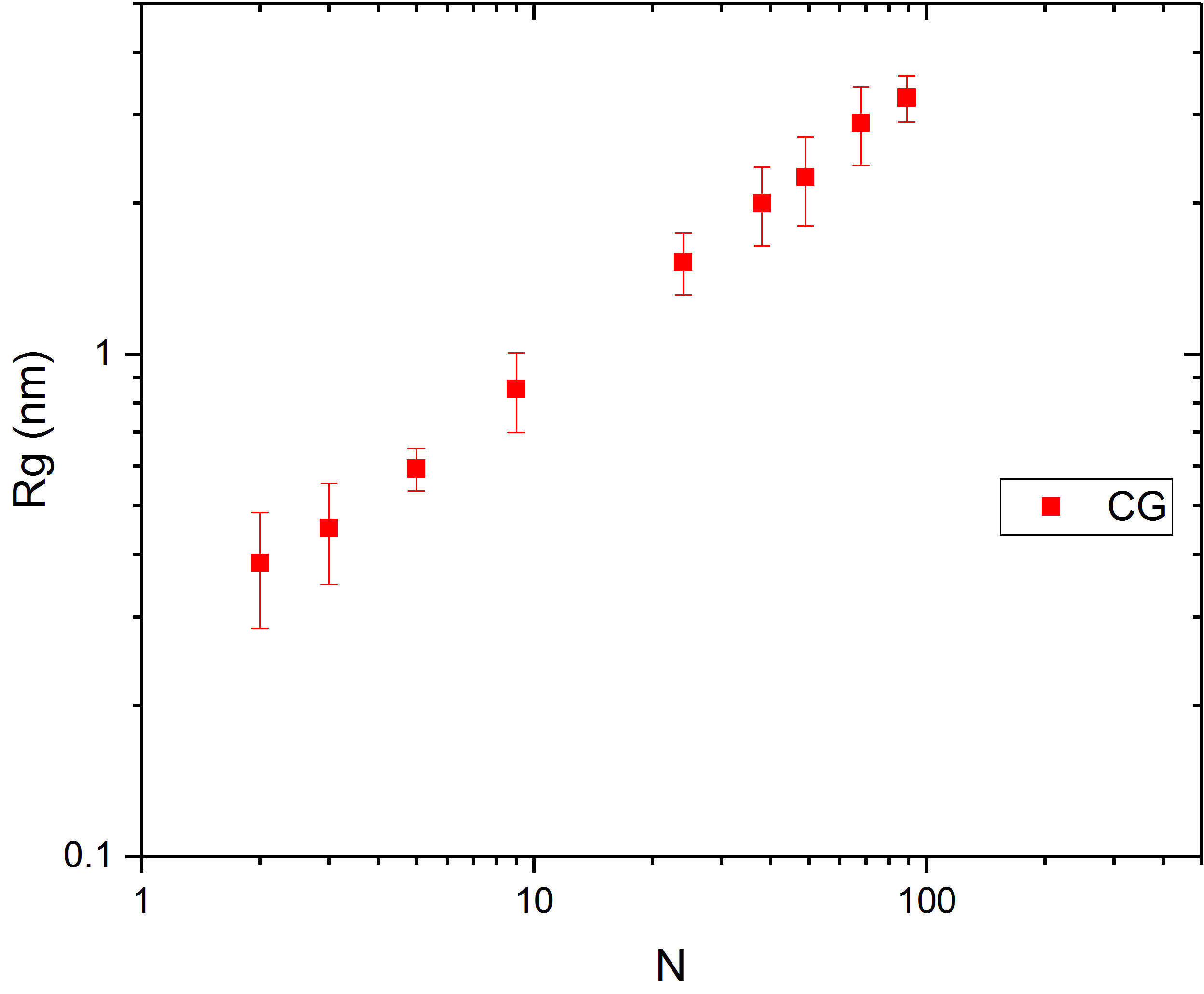}
\caption{$R_g$ of poly (N-(2-carboxyethyl) glycine) in
acetonitrile as a function of the repeat units number $N$ in CG simulation.}
\label{fig:Rg_b}
\end{figure}

Figure \ref{fig:Rg_b} presents $R_g$ as a function of $N$ of poly
(N-(2-carboxyethyl) glycine) in acetonitrile obtained with CG model. The fitted slope \(\nu\) is 0.583, which is close to the theoretical value $\nu=0.59$ for a peptoid in a good solvent. 
Note that \(\varepsilon_{P3 - SNd1}\)=
\SI{4.497}{KJ/mol} is smaller than \(\varepsilon_{P3 - P3}\)= \SI{5.0}{KJ/mol} in the MARTINI FF, where P3 denotes side-chain bead of poly (N-(2-carboxyethyl) glycine) and SNd1 stands for acetonitrile bead. 
This parameterization
would lead to the peptoid chain
collapse as a result of strong attraction between side chains, which makes acetonitrile behave as a poor solvent. Therefore, in the proposed CG model, we reduce
\(\varepsilon_{P3 - P3}\) to \SI{4.0}{KJ/mol}. Note that in practice, poly
(N-(2-carboxyethyl) glycine) side chains deprotonate and get solvated by
acetonitrile\cite{RN680}, keeping only a portion of hydroxyl
groups. Because of this, a very few hydrogen bonds are
formed between poly (N-(2-carboxyethyl) glycine) side chains. The
deprotonation cannot be accurately represented in classical MD simulation
that assumes a permanent chemical bond between hydrogen and oxygen atoms
in the hydroxyl group. As a result, in classical MD simulation it would predict more
hydrogen bonds between side chains so that the chain collapses.
Therefore, we do not use the atomic results as the reference for this system.

Figures  \ref{fig:Rg} and \ref{fig:Rg_b} demonstrate that in the proposed CG model, the bonded and
nonbonded potentials estimated for polysarcosine can be transferred to
the CG model of poly (N-(2-carboxyethyl) glycine) in good solvents. The
transferability of bond stretching and bond angle bending potentials to similar molecules
was also observed in other CG models.\cite{RN509} Here, we show
that the torsion potential also can be transferred in a properly
constructed CG model. Note that not all CG potentials of polymer model are
transferable. For example, in the study of CG polyethylene oxide (PEO)
model,\cite{RN284} nonbonded and bonded interactions derived
under different CG FFs could not be combined to accurately predict the
behavior of PEO chain in water.

\subsection{The effect of chain length on solvation free energy}

\begin{figure}
\includegraphics[width=3.13542in,height=2.58571in]{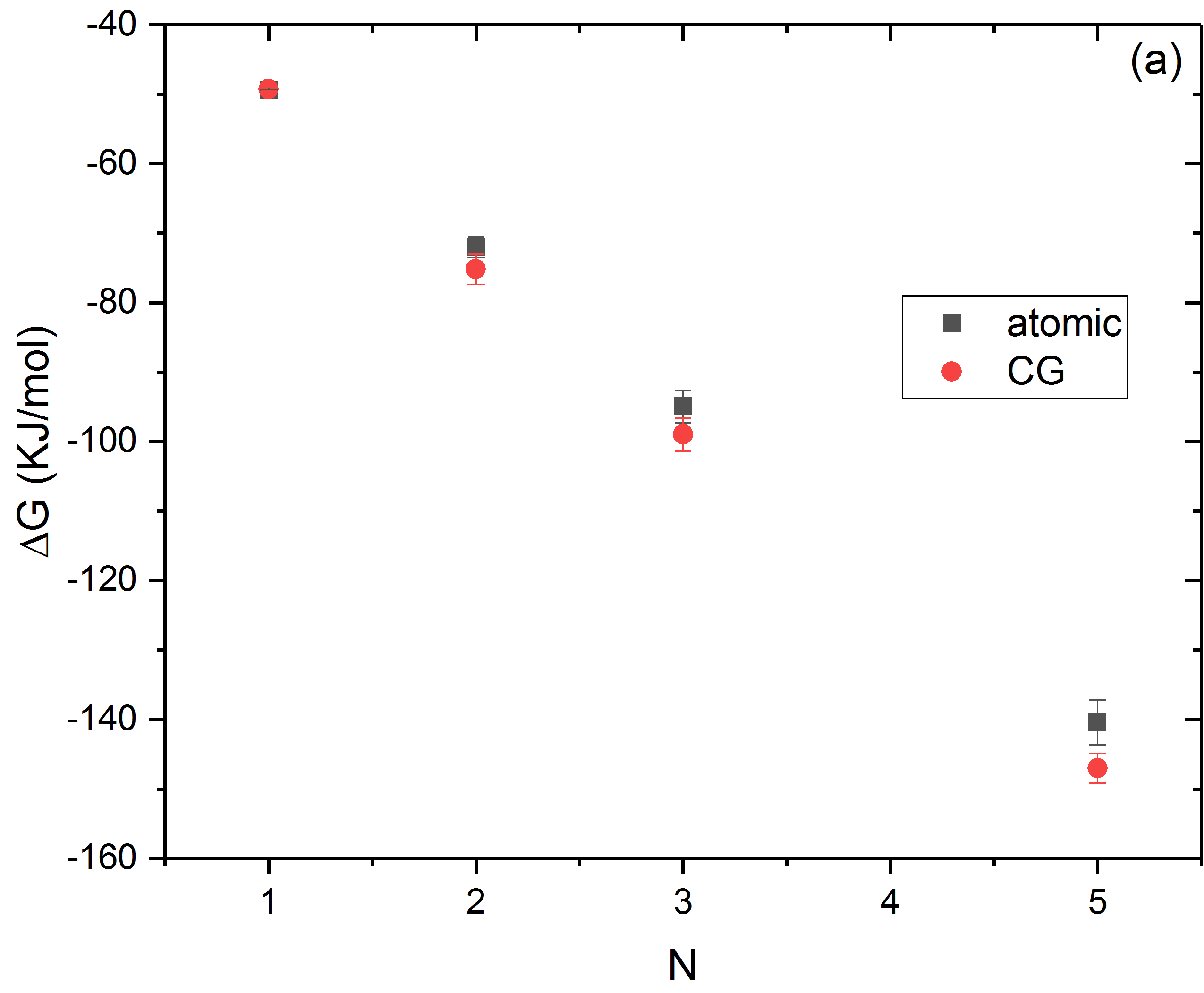}
\includegraphics[width=3.16166in,height=2.55162in]{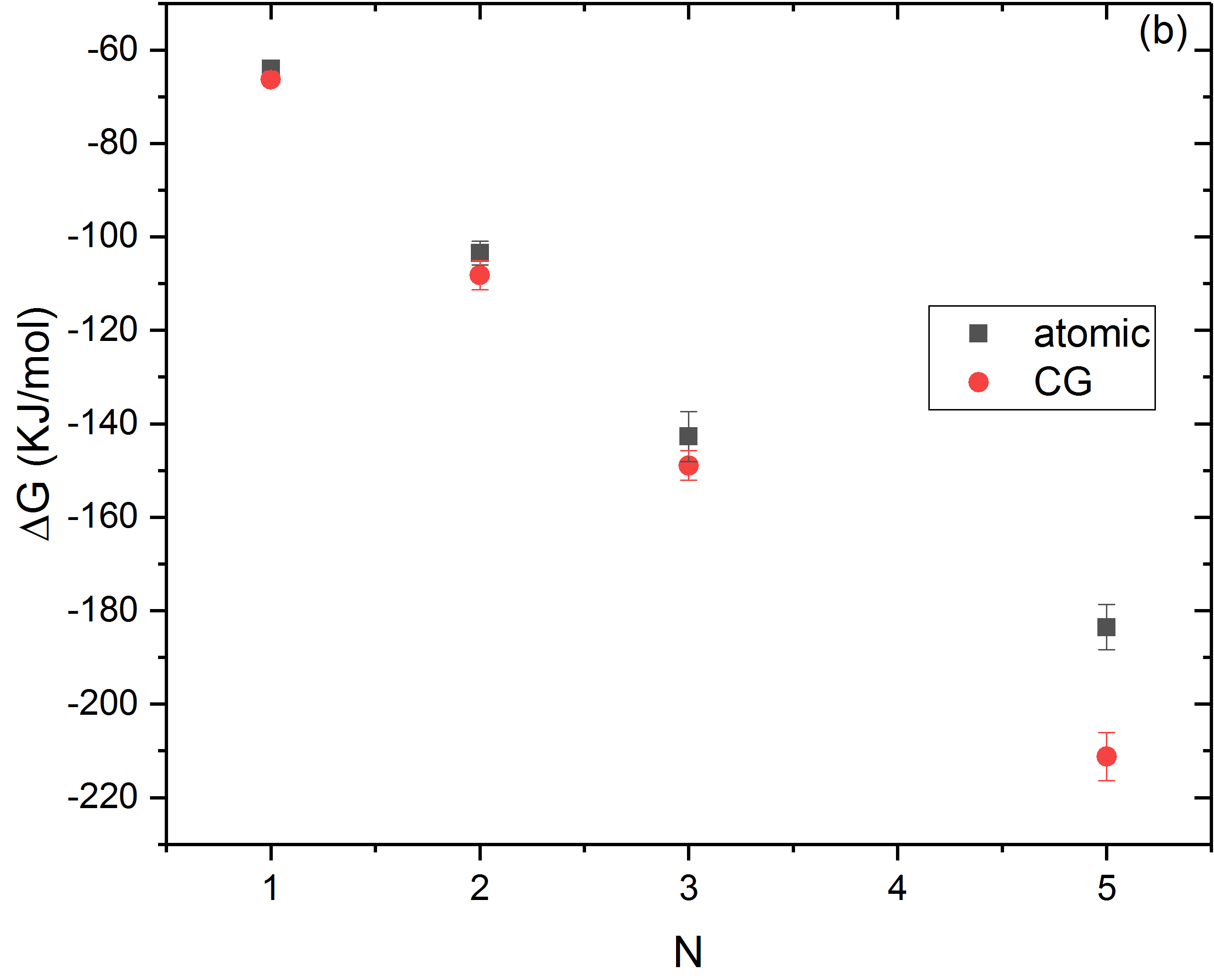}
\caption{(a)~Solvation free energy of polysarcosine in acetonitrile
with increasing number of repeat units. (b)~Solvation free energy of poly
(N-(2-carboxyethyl) glycine) in water with increasing number of repeat units.}
\label{Fig:nine}
\end{figure}

We investigate the effect of chain length on the solvation free energy
of polysarcosine and poly (N-(2-carboxyethyl) glycine). Figure \ref{Fig:nine} reports
the solvation free energy of polysarcosine in acetonitrile and 
poly (N-(2-carboxyethyl) glycine) in water as a function of the repeat unit number \(N\), obtained from CG and atomic simulations. We can see that in both atomic
and CG simulations, the solvation free energy increases with \(N\)
for both peptoids. Furthermore, we see that the CG model is transferable with respect to free energy, i.e., it can reproduce
the solvation free energy of polysarcosine in acetonitrile and poly
(N-(2-carboxyethyl) glycine) in water within 6\% (except for the poly
(N-(2-carboxyethyl) glycine) in water with $N=5$ where the error is about 16\%).
A possible reason for the weaker transferability of the CG model for poly
(N-(2-carboxyethyl) glycine) in water is that it does not take into account the effect of the hydrogen bonding between the hydroxyl groups. 
In the atomic model of
poly (N-(2-carboxyethyl) glycine) in water, the hydrogen bonding can be
formed between the hydroxyl groups on the side chain, which affects the
conformation of the backbone. Our atomic simulations show that as the
chain length increases, the poly (N-(2-carboxyethyl) glycine) chain tends
to collapse in water. Since the CG torsion potential is obtained from a
polysarcosine in a good solvent (acetonitrile), it predicts a more
extended conformation of poly (N-(2-carboxyethyl) glycine) in water.
%Therefore, the transferability of backbone torsion potential for long
%poly (N-(2-carboxyethyl) glycine) chain in water will be weak as a result
%of the special interaction between side chains. 
Because of the high computational cost of the BAR estimate of \(\Delta G\), we limit our study the maximum chain length
\(N = 5\) in atomic simulations.
%This method
%requires conducting at least 20 simulations for each chain length.
%Additional study is needed to further test the transferability
%of CG model for longer peptoid chains.

\subsection {Coil-to-globule transition of polypeptoid chain}
Polymer collapse is the simplest form of protein folding, which is
caused by the intramolecular interactions and solvent entropy. In this
section, we use our CG model to study unfolding of an initially coiled
polysarcosine/poly (N-(2-carboxyethyl) glycine) diblock polypeptoid chain in
water-acetonitrile mixture. The choice of this peptoid is motivated by
the experiments\cite{RN207} 
on the coil-to-globule transition, where the hydrophobic interactions are concluded to be the major driving force of the peptoid chain collapse.
\begin{figure}
\includegraphics[scale=0.5]{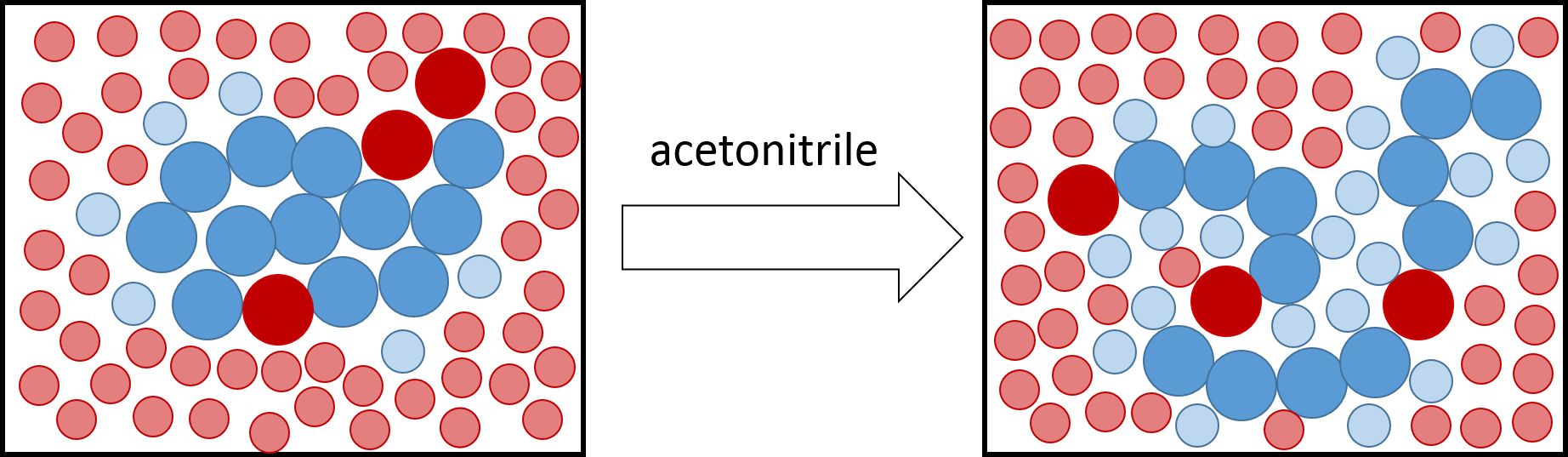}
\caption{The scheme of coil-to-globule transition for diblock peptoid
chain. Red circles represent N-(2-carboxyethyl)
glycine monomers, blue circle are sarcosine monomers, light red and
light blue circles are water and acetonitrile, respectevely.}
\label{fig:fig9}
\end{figure}
%
%\begin{figure}
%\includegraphics[scale=0.35]{media/10new.png}
%\caption{$R_g$ of diblock peptoid chain versus the acetonitrile concentration in experiment (red) and CG simulation (black).  Experimental data are from Ref. 65.}
%\label{fig:fig10}
%\end{figure}
%
\begin{figure}
\includegraphics[scale=0.8]{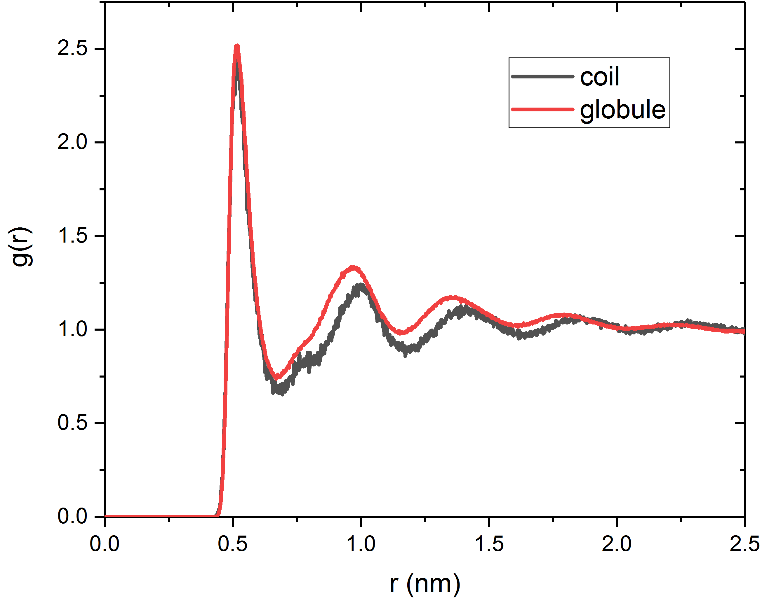}
\caption{Radial distribution function of the diblock poly peptoid in acetonitrile for coil and stretch configurations obtained from the CG model.}
\label{fig:s5}
\end{figure}
We simulate the behavior of the sequenced polysarcosine/poly
(N-(2-carboxyethyl) glycine) diblock peptoid in a water-acetonitrile
mixture for different acetonitrile concentrations. The sequenced polysarcosine/poly
(N-(2-carboxyethyl) glycine) diblock peptoid chain includes 100~CG beads
and the repeat unit is one (N-(2-carboxyethyl) glycine) bead and four
sarcosine beads. In the experiment,\cite{RN207} the polypeptoid chain
were found to be coiled at low concentrations of acetonitrile and swell and form a globule at higher concentrations, as
schematically shown in Figure \ref{fig:fig9}. 
Our CG model predicts the $R_g$ in the coil state of $2.15\pm 0.19${nm}, which is close to the $R_g$ values of 2.2 nm observed in the experiment\cite{RN207}. In the globule state, our model predicts $R_g=3.4\pm 0.25$nm, which is close the experimental value of $3.5$.
 
Figure \ref{fig:s5} shows the peptoid RDF for the lowest and highest considered acetonitrile concentrations.  
The RDF peaks in the higher concentration mixture are higher than those in the lower concentration mixture. The increase in both RDF peaks and $R_g$ with the acetonitrile concentrations indicates that peptoids is swollen as the acetonitrile concentration increases.  

\begin{comment}
In Figure  \ref{fig:fig10}, the transition from lower to upper asymptotic $R_g$ values reflect the transition of peptoid from coiled to globule states, and $R_g$ of the peptoid in these two states predicted by our CG model is in good agreement with the experimental observations.  
In the experiment, the coil-to-globule transition occurs at the concentration of acetonitrile of approximately 11 M while in our simulations the transition happens at the concentration of approximately \SI{75}{M}. One reason for this is that the MARTINI FF for water, used in our CG model, underestimates the absolute self-solvation free energy. This may imply stronger attraction between peptoid and water than in the real system. To complete the coil-to-globule transition, the polypeptoid chain needs to 
be solvated by acetonitrile. Therefore, higher concentration of acetonitrile is needed to solvate a polypeptoid chain in our CG model. The prediction of transition concentration can be improved by optimizing water potential with respect to the solvation free energy in the MARTINI FF. 
\end{comment}

\section {Conclusion}

In this paper, we proposed a new CG model for poly ($\alpha$-peptoid)s.   We built
the CG model for the poly ($\alpha$-peptoid) backbone (polysarcosine) in various solvents, including water, hexane, 1-octanol, and acetonitrile,
and extended it to other peptoids. 
In our CG polypeptoid solution, we had three types of beads, including backbone, sidechain, and solvent beads, and the same degree of coarse-graining at in the MARTINI FF. This makes our model compatible with the MARTINI FF.  
%In our model,
%one CG bead of the backbone represents a repeat unit of the backbone's atomic structure.  
% As a result, the CG model of backbone is made of identical beads and
%each side chain is represented by a CG bead composed of four heavy atoms.
All interactions between beads were divided into bonded interaction and
nonbonded interaction. The bonded interactions between CG beads were
parameterized to reproduce local and global structural properties
including PDFs of bond length, bond angle, and dihedral angle and gyration radius obtained
from atomic simulations of polypeptoids. 
%Nonbonded interactions are parameterized to reproduce 
%transfer free energy and absolute solvation free energy
Nonbonded potentials for water, hexane, and 1-octanol as well as potential for the interaction between hydrophilic or hydrophobic side chains and these solvents were taken from the MARTINI FF. The potential for acetonitrile and potentials between solvents and the backbone which are not given in MARTINI FF were parameterized by matching not only the transfer free energy but also the absolute solvation free energy. Our parameterization approach is expected to be more accurate than the approach in MARTINI FF where the target property is only the transfer free energy. The CG parameters of the 
backbone were extended to other peptoids with hydrophobic or hydrophilic
side chains to examine transferability of the proposed CG model. 
%We demonstrated that the nonbonded interactions between side chains and solvents could be built by simply adding side chain parameters taken from the MARTINI FF, respectively. 
We found that the hydration free
energy of poly
(N-(2-carboxyethyl) glycine) and poly (N-pentyl glycine) peptoids, computed from atomic simulations, is well reproduced by our CG model combining the proposed bonded potentials and nonbonded potentials with MARTINI FF.
 These results suggest that a CG model of any poly
($\alpha$-peptoid) can be constructed by adding side chains to our polysarcosine CG model with the nonbonded potentials given by MARTINI FF.

%can be extended viewed a generic model for poly
%$\alpha$-peptoids, as any poly $\alpha$-peptoid can be modeled with the proposed model
%by adding side groups with parameters from the Martini CG FF.

We evaluated the transferability of the bonded interaction parameters
(bond, angle and torsion potentials) in the CG polysarcosine model
by comparison of the local conformational PDFs of polysarcosine in various
solvents. The previous CG models \cite{RN481,RN2} found that the bond and angle potentials to be transferable and torsion potential to be non-transferable in solution. In our CG model, the bond and angle potentials for polysarcosine are transferable for all considered solvents and the torsion potential is only transferable for good solvent.  We demonstrated the transferability of torsion potential in good solvent with respect to the chain length by comparing the Rg of polysarcosine in a good solvent at modestly high molecular weight in atomic and CG simulations. We also found good transferability of the CG backbone parameter to poly (N-(2-carboxyethyl) glycine) in good solvent with respect to the chain length.
%
%To study the limitations on transferability of the torsion potential, we
%considered the effect of the intramolecular nonbonded interactions between beads representing two sidechains of the same peptoid molecule on the torsion of the peptoid backbone. Specifically,  we compared two poly $\alpha$-peptoids with
%or without sidechains in the same good solvent. We found that the CG torsion potential is transferable in the absence of special
%interactions (i.e., hydrogen bonding) between sidechains. 
%up to 90 repeated units. 
Next, we demonstrated that the nonbonded potentials are transferable with respect to the solvation free energy for peptoids oligomers
with the backbone made of five or less repeat units. We calculated the solvation free energy
of two different polypeptoids, polysarcosine in acetonitrile and poly
(N-(2-carboxyethyl) glycine) in water. The difference between the free energy in CG and atomic predictions for polysarcosine in acetonitrile is less than 6\% for all considered chain lengths. The error in the solvation free energy 
of poly (N-(2-carboxyethyl) glycine) in water is less than 6\% for chain length less than four repeat units and increases to about 16\% for the five-unit-long peptoid chain. Note that the hydrogen bonding between side chains makes water like a poor solvent for poly (N-(2-carboxyethyl) glycine) in the atomic simulation here.
Given that the
torsion potential is not used for
poly ($\alpha$-peptoid) backbone length less than four CG beads, the relatively large error for the peptoid with backbone larger than four beads indicate that 
the torsion potential has weak transferable in poor solvents. On the other hand, the simulations of polysarcosine in acetonitrile confirmed that the torsion potential is transferable in good solvents. 
Finally, we demonstrated that our CG model can describe the coil-to-globule transition of diblock polypeptoid chain in water-acetonitrile mixture and accurately predict the radius of gyration at both coil and globule states. 
%We also observed that our model overestimates the transition concentration. This could be due to the nonbonded water-water interactions in MARTINI FF (used in our CG model) that underestimates the self-solvation free energy. Therefore, the prediction of the transition concentration can be improved by optimizing parameters in this potential with respect to the solvation free energy. 

In this work, we demonstrated transferability of the CG potentials with respect to solvation free energy and conformation state. However, transferability of CG polymer model in solution with respect to other thermodynamic conditions remains an open question. For example, most of existing CG FF, including MARTINI FF, cannot describe phase transition (e.g.,
crystallization\cite{RN686}) of biopolymer because of the fixed backbone structure and complex interactions of the side chains. More advanced methods (e.g., machine learning) may be needed to
generate adaptive CG potentials to describe the phase transition of polypeptoid solutions. 
%In this case, nonbonded
%interactions must be reoptimized from crystal structure to correct
%the contribution of lattice energy in CG potential. 

\begin{acknowledgement}
This work was supported by the U.S. Department of Energy (DOE),
Office of Science, Office of Advanced Scientific Computing
Research. Pacific Northwest National Laboratory is operated
by Battelle for DOE under Contract DE-AC05-76RL01830.
\end{acknowledgement}

\suppinfo
The Supporting Information is available free of charge on the
ACS Publications website.

\bibliography{peptoid}

\providecommand{\latin}[1]{#1}
\makeatletter
\providecommand{\doi}
  {\begingroup\let\do\@makeother\dospecials
  \catcode`\{=1 \catcode`\}=2 \doi@aux}
\providecommand{\doi@aux}[1]{\endgroup\texttt{#1}}
\makeatother
\providecommand*\mcitethebibliography{\thebibliography}
\csname @ifundefined\endcsname{endmcitethebibliography}
  {\let\endmcitethebibliography\endthebibliography}{}
\begin{mcitethebibliography}{70}
\providecommand*\natexlab[1]{#1}
\providecommand*\mciteSetBstSublistMode[1]{}
\providecommand*\mciteSetBstMaxWidthForm[2]{}
\providecommand*\mciteBstWouldAddEndPuncttrue
  {\def\EndOfBibitem{\unskip.}}
\providecommand*\mciteBstWouldAddEndPunctfalse
  {\let\EndOfBibitem\relax}
\providecommand*\mciteSetBstMidEndSepPunct[3]{}
\providecommand*\mciteSetBstSublistLabelBeginEnd[3]{}
\providecommand*\EndOfBibitem{}
\mciteSetBstSublistMode{f}
\mciteSetBstMaxWidthForm{subitem}{(\alph{mcitesubitemcount})}
\mciteSetBstSublistLabelBeginEnd
  {\mcitemaxwidthsubitemform\space}
  {\relax}
  {\relax}

\bibitem[Robertson \latin{et~al.}(2016)Robertson, Battigelli, Proulx, Mannige,
  Haxton, Yun, Whitelam, and Zuckermann]{RN70}
Robertson,~E.~J.; Battigelli,~A.; Proulx,~C.; Mannige,~R.~V.; Haxton,~T.~K.;
  Yun,~L.; Whitelam,~S.; Zuckermann,~R.~N. Design, Synthesis, Assembly, and
  Engineering of Peptoid Nanosheets. \emph{Acc. Chem. Res.} \textbf{2016},
  \emph{49}, 379--389\relax
\mciteBstWouldAddEndPuncttrue
\mciteSetBstMidEndSepPunct{\mcitedefaultmidpunct}
{\mcitedefaultendpunct}{\mcitedefaultseppunct}\relax
\EndOfBibitem
\bibitem[Knight \latin{et~al.}(2017)Knight, Kulkarni, Zhou, Franke, Miller, and
  Francis]{RN275}
Knight,~A.~S.; Kulkarni,~R.~U.; Zhou,~E.~Y.; Franke,~J.~M.; Miller,~E.~W.;
  Francis,~M.~B. A modular platform to develop peptoid-based selective
  fluorescent metal sensors. \emph{Chem. Commun.} \textbf{2017}, \emph{53},
  3477--3480\relax
\mciteBstWouldAddEndPuncttrue
\mciteSetBstMidEndSepPunct{\mcitedefaultmidpunct}
{\mcitedefaultendpunct}{\mcitedefaultseppunct}\relax
\EndOfBibitem
\bibitem[Lau(2014)]{RN343}
Lau,~K. H.~A. Peptoids for biomaterials science. \emph{Biomater. Sci.}
  \textbf{2014}, \emph{2}, 627--633\relax
\mciteBstWouldAddEndPuncttrue
\mciteSetBstMidEndSepPunct{\mcitedefaultmidpunct}
{\mcitedefaultendpunct}{\mcitedefaultseppunct}\relax
\EndOfBibitem
\bibitem[Jun \latin{et~al.}(2015)Jun, Altoe, Aloni, and Zuckermann]{RN386}
Jun,~J. M.~V.; Altoe,~M. V.~P.; Aloni,~S.; Zuckermann,~R.~N. Peptoid nanosheets
  as soluble, two-dimensional templates for calcium carbonate mineralization.
  \emph{Chem. Commun.} \textbf{2015}, \emph{51}, 10218--10221\relax
\mciteBstWouldAddEndPuncttrue
\mciteSetBstMidEndSepPunct{\mcitedefaultmidpunct}
{\mcitedefaultendpunct}{\mcitedefaultseppunct}\relax
\EndOfBibitem
\bibitem[Merrill \latin{et~al.}(2018)Merrill, Yan, Jin, Mu, Chen, and
  Knecht]{RN685}
Merrill,~N.~A.; Yan,~F.; Jin,~H.; Mu,~P.; Chen,~C.-L.; Knecht,~M.~R. Tunable
  assembly of biomimetic peptoids as templates to control nanostructure
  catalytic activity. \emph{Nanoscale} \textbf{2018}, \emph{10},
  12445--12452\relax
\mciteBstWouldAddEndPuncttrue
\mciteSetBstMidEndSepPunct{\mcitedefaultmidpunct}
{\mcitedefaultendpunct}{\mcitedefaultseppunct}\relax
\EndOfBibitem
\bibitem[Patterson \latin{et~al.}(2017)Patterson, Wenning, Rizis, Calabrese,
  Finlay, Franco, Zuckermann, Clare, Kramer, Ober, and Segalman]{RN277}
Patterson,~A.~L.; Wenning,~B.; Rizis,~G.; Calabrese,~D.~R.; Finlay,~J.~A.;
  Franco,~S.~C.; Zuckermann,~R.~N.; Clare,~A.~S.; Kramer,~E.~J.; Ober,~C.~K.
  \latin{et~al.}  Role of Backbone Chemistry and Monomer Sequence in
  Amphiphilic Oligopeptide- and Oligopeptoid-Functionalized PDMS- and PEO-Based
  Block Copolymers for Marine Antifouling and Fouling Release Coatings.
  \emph{Macromolecules} \textbf{2017}, \emph{50}, 2656--2667\relax
\mciteBstWouldAddEndPuncttrue
\mciteSetBstMidEndSepPunct{\mcitedefaultmidpunct}
{\mcitedefaultendpunct}{\mcitedefaultseppunct}\relax
\EndOfBibitem
\bibitem[Reyes \latin{et~al.}(2014)Reyes, Guo, Hedgepeth, Zhang, and
  Kelland]{RN125}
Reyes,~F.~T.; Guo,~L.; Hedgepeth,~J.~W.; Zhang,~D.; Kelland,~M.~A. First
  Investigation of the Kinetic Hydrate Inhibitor Performance of
  Poly(N-alkylglycine)s. \emph{Energy and Fuels} \textbf{2014}, \emph{28},
  6889--6896\relax
\mciteBstWouldAddEndPuncttrue
\mciteSetBstMidEndSepPunct{\mcitedefaultmidpunct}
{\mcitedefaultendpunct}{\mcitedefaultseppunct}\relax
\EndOfBibitem
\bibitem[Knight \latin{et~al.}(2015)Knight, Zhou, Francis, and
  Zuckermann]{RN182}
Knight,~A.~S.; Zhou,~E.~Y.; Francis,~M.~B.; Zuckermann,~R.~N. Sequence
  Programmable Peptoid Polymers for Diverse Materials Applications. \emph{Adv.
  Mater.} \textbf{2015}, \emph{27}, 5665--5691\relax
\mciteBstWouldAddEndPuncttrue
\mciteSetBstMidEndSepPunct{\mcitedefaultmidpunct}
{\mcitedefaultendpunct}{\mcitedefaultseppunct}\relax
\EndOfBibitem
\bibitem[Weber \latin{et~al.}(2018)Weber, Birke, Fischer, Schmidt, and
  Barz]{RN664}
Weber,~B.; Birke,~A.; Fischer,~K.; Schmidt,~M.; Barz,~M. Solution Properties of
  Polysarcosine: From Absolute and Relative Molar Mass Determinations to
  Complement Activation. \emph{Macromolecules} \textbf{2018}, \emph{51},
  2653--2661\relax
\mciteBstWouldAddEndPuncttrue
\mciteSetBstMidEndSepPunct{\mcitedefaultmidpunct}
{\mcitedefaultendpunct}{\mcitedefaultseppunct}\relax
\EndOfBibitem
\bibitem[Hara \latin{et~al.}(2014)Hara, Ueda, Makino, Hara, Ozeki, and
  Kimura]{RN387}
Hara,~E.; Ueda,~M.; Makino,~A.; Hara,~I.; Ozeki,~E.; Kimura,~S. Factors
  Influencing in Vivo Disposition of Polymeric Micelles on Multiple
  Administrations. \emph{ACS Med. Chem. Lett.} \textbf{2014}, \emph{5},
  873--877\relax
\mciteBstWouldAddEndPuncttrue
\mciteSetBstMidEndSepPunct{\mcitedefaultmidpunct}
{\mcitedefaultendpunct}{\mcitedefaultseppunct}\relax
\EndOfBibitem
\bibitem[Ueda \latin{et~al.}(2011)Ueda, Makino, Imai, Sugiyama, and
  Kimura]{RN388}
Ueda,~M.; Makino,~A.; Imai,~T.; Sugiyama,~J.; Kimura,~S. Temperature-Triggered
  Fusion of Vesicles Composed of Right-Handed and Left-Handed Amphiphilic
  Helical Peptides. \emph{Langmuir} \textbf{2011}, \emph{27}, 4300--4304\relax
\mciteBstWouldAddEndPuncttrue
\mciteSetBstMidEndSepPunct{\mcitedefaultmidpunct}
{\mcitedefaultendpunct}{\mcitedefaultseppunct}\relax
\EndOfBibitem
\bibitem[Sano \latin{et~al.}(2017)Sano, Ohashi, Kanazaki, Makino, Ding,
  Deguchi, Kanada, Ono, and Saji]{RN389}
Sano,~K.; Ohashi,~M.; Kanazaki,~K.; Makino,~A.; Ding,~N.; Deguchi,~J.;
  Kanada,~Y.; Ono,~M.; Saji,~H. Indocyanine Green-Labeled Polysarcosine for in
  Vivo Photoacoustic Tumor Imaging. \emph{Bioconjug. Chem.} \textbf{2017},
  \emph{28}, 1024--1030\relax
\mciteBstWouldAddEndPuncttrue
\mciteSetBstMidEndSepPunct{\mcitedefaultmidpunct}
{\mcitedefaultendpunct}{\mcitedefaultseppunct}\relax
\EndOfBibitem
\bibitem[Zhu \latin{et~al.}(2017)Zhu, Chen, Yan, Chen, Tao, Ling, Yang, He, and
  Mao]{RN348}
Zhu,~H.; Chen,~Y.; Yan,~F.-J.; Chen,~J.; Tao,~X.-F.; Ling,~J.; Yang,~B.;
  He,~Q.-J.; Mao,~Z.-W. Polysarcosine brush stabilized gold nanorods for in
  vivo near-infrared photothermal tumor therapy. \emph{Acta Biomater.}
  \textbf{2017}, \emph{50}, 534--545\relax
\mciteBstWouldAddEndPuncttrue
\mciteSetBstMidEndSepPunct{\mcitedefaultmidpunct}
{\mcitedefaultendpunct}{\mcitedefaultseppunct}\relax
\EndOfBibitem
\bibitem[Luxenhofer \latin{et~al.}(2013)Luxenhofer, Fetsch, and
  Grossmann]{RN122}
Luxenhofer,~R.; Fetsch,~C.; Grossmann,~A. Polypeptoids: A perfect match for
  molecular definition and macromolecular engineering? \emph{J. Polym. Sci.,
  Part A: Polym. Chem.} \textbf{2013}, \emph{51}, 2731--2752\relax
\mciteBstWouldAddEndPuncttrue
\mciteSetBstMidEndSepPunct{\mcitedefaultmidpunct}
{\mcitedefaultendpunct}{\mcitedefaultseppunct}\relax
\EndOfBibitem
\bibitem[Douy and Gallot(1987)Douy, and Gallot]{RN394}
Douy,~A.; Gallot,~B. Amphipathic block copolymers with two polypeptide blocks:
  Synthesis and structural study of
  poly(N-trifluoroacetyl-l-lysine)-polysarcosine copolymers. \emph{Polymer}
  \textbf{1987}, \emph{28}, 147--154\relax
\mciteBstWouldAddEndPuncttrue
\mciteSetBstMidEndSepPunct{\mcitedefaultmidpunct}
{\mcitedefaultendpunct}{\mcitedefaultseppunct}\relax
\EndOfBibitem
\bibitem[Kimura and Imanishi(1983)Kimura, and Imanishi]{RN396}
Kimura,~S.; Imanishi,~Y. Synthesis and conformation of the cyclic octapeptides
  cyclo(Phe-Pro)4, cyclo(Leu-Pro)4, and cyclo[Lys(Z)-Pro]4. \emph{Biopolymers}
  \textbf{1983}, \emph{22}, 2191--2206\relax
\mciteBstWouldAddEndPuncttrue
\mciteSetBstMidEndSepPunct{\mcitedefaultmidpunct}
{\mcitedefaultendpunct}{\mcitedefaultseppunct}\relax
\EndOfBibitem
\bibitem[Fokina \latin{et~al.}(2016)Fokina, Klinker, Braun, Jeong, Bae, Barz,
  and Zentel]{RN378}
Fokina,~A.; Klinker,~K.; Braun,~L.; Jeong,~B.~G.; Bae,~W.~K.; Barz,~M.;
  Zentel,~R. Multidentate Polysarcosine-Based Ligands for Water-Soluble Quantum
  Dots. \emph{Macromolecules} \textbf{2016}, \emph{49}, 3663--3671\relax
\mciteBstWouldAddEndPuncttrue
\mciteSetBstMidEndSepPunct{\mcitedefaultmidpunct}
{\mcitedefaultendpunct}{\mcitedefaultseppunct}\relax
\EndOfBibitem
\bibitem[Klinker and Barz(2015)Klinker, and Barz]{RN393}
Klinker,~K.; Barz,~M. Polypept(o)ides: Hybrid Systems Based on Polypeptides and
  Polypeptoids. \emph{Macromol. Rapid Commun.} \textbf{2015}, \emph{36},
  1943--1957\relax
\mciteBstWouldAddEndPuncttrue
\mciteSetBstMidEndSepPunct{\mcitedefaultmidpunct}
{\mcitedefaultendpunct}{\mcitedefaultseppunct}\relax
\EndOfBibitem
\bibitem[Hortz \latin{et~al.}(2015)Hortz, Birke, Kaps, Decker, Wächtersbach,
  Fischer, Schuppan, Barz, and Schmidt]{RN286}
Hortz,~C.; Birke,~A.; Kaps,~L.; Decker,~S.; Wächtersbach,~E.; Fischer,~K.;
  Schuppan,~D.; Barz,~M.; Schmidt,~M. Cylindrical Brush Polymers with
  Polysarcosine Side Chains: A Novel Biocompatible Carrier for Biomedical
  Applications. \emph{Macromolecules} \textbf{2015}, \emph{48},
  2074--2086\relax
\mciteBstWouldAddEndPuncttrue
\mciteSetBstMidEndSepPunct{\mcitedefaultmidpunct}
{\mcitedefaultendpunct}{\mcitedefaultseppunct}\relax
\EndOfBibitem
\bibitem[Birke \latin{et~al.}(2014)Birke, Huesmann, Kelsch, Weilbächer, Xie,
  Bros, Bopp, Becker, Landfester, and Barz]{RN392}
Birke,~A.; Huesmann,~D.; Kelsch,~A.; Weilbächer,~M.; Xie,~J.; Bros,~M.;
  Bopp,~T.; Becker,~C.; Landfester,~K.; Barz,~M. Polypeptoid-block-polypeptide
  Copolymers: Synthesis, Characterization, and Application of Amphiphilic Block
  Copolypept(o)ides in Drug Formulations and Miniemulsion Techniques.
  \emph{Biomacromolecules} \textbf{2014}, \emph{15}, 548--557\relax
\mciteBstWouldAddEndPuncttrue
\mciteSetBstMidEndSepPunct{\mcitedefaultmidpunct}
{\mcitedefaultendpunct}{\mcitedefaultseppunct}\relax
\EndOfBibitem
\bibitem[Park and Szleifer(2011)Park, and Szleifer]{RN40}
Park,~S.~H.; Szleifer,~I. Structural and Dynamical Characteristics of Peptoid
  Oligomers with Achiral Aliphatic Side Chains Studied by Molecular Dynamics
  Simulation. \emph{J. Phys. Chem. B} \textbf{2011}, \emph{115},
  10967--10975\relax
\mciteBstWouldAddEndPuncttrue
\mciteSetBstMidEndSepPunct{\mcitedefaultmidpunct}
{\mcitedefaultendpunct}{\mcitedefaultseppunct}\relax
\EndOfBibitem
\bibitem[Mannige \latin{et~al.}(2015)Mannige, Haxton, Proulx, Robertson,
  Battigelli, Butterfoss, Zuckermann, and Whitelam]{RN402}
Mannige,~R.~V.; Haxton,~T.~K.; Proulx,~C.; Robertson,~E.~J.; Battigelli,~A.;
  Butterfoss,~G.~L.; Zuckermann,~R.~N.; Whitelam,~S. Peptoid nanosheets exhibit
  a new secondary-structure motif. \emph{Nature} \textbf{2015}, \emph{526},
  415--420\relax
\mciteBstWouldAddEndPuncttrue
\mciteSetBstMidEndSepPunct{\mcitedefaultmidpunct}
{\mcitedefaultendpunct}{\mcitedefaultseppunct}\relax
\EndOfBibitem
\bibitem[Daily \latin{et~al.}(2016)Daily, Baer, and Mundy]{RN400}
Daily,~M.~D.; Baer,~M.~D.; Mundy,~C.~J. Divalent Ion Parameterization Strongly
  Affects Conformation and Interactions of an Anionic Biomimetic Polymer.
  \emph{J. Phys. Chem. B} \textbf{2016}, \emph{120}, 2198--2208\relax
\mciteBstWouldAddEndPuncttrue
\mciteSetBstMidEndSepPunct{\mcitedefaultmidpunct}
{\mcitedefaultendpunct}{\mcitedefaultseppunct}\relax
\EndOfBibitem
\bibitem[Darré \latin{et~al.}(2015)Darré, Machado, Brandner, González,
  Ferreira, and Pantano]{RN521}
Darré,~L.; Machado,~M.~R.; Brandner,~A.~F.; González,~H.~C.; Ferreira,~S.;
  Pantano,~S. SIRAH: A Structurally Unbiased Coarse-Grained Force Field for
  Proteins with Aqueous Solvation and Long-Range Electrostatics. \emph{J. Chem.
  Theory Comput.} \textbf{2015}, \emph{11}, 723--739\relax
\mciteBstWouldAddEndPuncttrue
\mciteSetBstMidEndSepPunct{\mcitedefaultmidpunct}
{\mcitedefaultendpunct}{\mcitedefaultseppunct}\relax
\EndOfBibitem
\bibitem[Mukherji \latin{et~al.}(2014)Mukherji, Marques, and Kremer]{RN634}
Mukherji,~D.; Marques,~C.~M.; Kremer,~K. Polymer collapse in miscible good
  solvents is a generic phenomenon driven by preferential adsorption.
  \emph{Nat. Commun.} \textbf{2014}, \emph{5}, 4882\relax
\mciteBstWouldAddEndPuncttrue
\mciteSetBstMidEndSepPunct{\mcitedefaultmidpunct}
{\mcitedefaultendpunct}{\mcitedefaultseppunct}\relax
\EndOfBibitem
\bibitem[Agrawal \latin{et~al.}(2014)Agrawal, Aryal, Perahia, Ge, and
  Grest]{RN528}
Agrawal,~A.; Aryal,~D.; Perahia,~D.; Ge,~T.; Grest,~G.~S. Coarse-Graining
  Atactic Polystyrene and Its Analogues. \emph{Macromolecules} \textbf{2014},
  \emph{47}, 3210--3218\relax
\mciteBstWouldAddEndPuncttrue
\mciteSetBstMidEndSepPunct{\mcitedefaultmidpunct}
{\mcitedefaultendpunct}{\mcitedefaultseppunct}\relax
\EndOfBibitem
\bibitem[Mantha and Yethiraj(2015)Mantha, and Yethiraj]{RN167}
Mantha,~S.; Yethiraj,~A. Conformational Properties of Sodium
  Polystyrenesulfonate in Water: Insights from a Coarse-Grained Model with
  Explicit Solvent. \emph{J. Phys. Chem. B} \textbf{2015}, \emph{119},
  11010--11018\relax
\mciteBstWouldAddEndPuncttrue
\mciteSetBstMidEndSepPunct{\mcitedefaultmidpunct}
{\mcitedefaultendpunct}{\mcitedefaultseppunct}\relax
\EndOfBibitem
\bibitem[Ozgur and Sayar(2016)Ozgur, and Sayar]{RN226}
Ozgur,~B.; Sayar,~M. Assembly of Triblock Amphiphilic Peptides into
  One-Dimensional Aggregates and Network Formation. \emph{J. Phys. Chem. B}
  \textbf{2016}, \emph{120}, 10243--10257\relax
\mciteBstWouldAddEndPuncttrue
\mciteSetBstMidEndSepPunct{\mcitedefaultmidpunct}
{\mcitedefaultendpunct}{\mcitedefaultseppunct}\relax
\EndOfBibitem
\bibitem[de~Oliveira \latin{et~al.}(2016)de~Oliveira, Netz, Kremer, Junghans,
  and Mukherji]{RN3}
de~Oliveira,~T.~E.; Netz,~P.~A.; Kremer,~K.; Junghans,~C.; Mukherji,~D. C-IBI:
  Targeting cumulative coordination within an iterative protocol to derive
  coarse-grained models of (multi-component) complex fluids. \emph{J. Chem.
  Phys.} \textbf{2016}, \emph{144}, 174106\relax
\mciteBstWouldAddEndPuncttrue
\mciteSetBstMidEndSepPunct{\mcitedefaultmidpunct}
{\mcitedefaultendpunct}{\mcitedefaultseppunct}\relax
\EndOfBibitem
\bibitem[Zhang and Guo(2014)Zhang, and Guo]{RN509}
Zhang,~J.; Guo,~H. Transferability of Coarse-Grained Force Field for nCB Liquid
  Crystal Systems. \emph{J. Phys. Chem. B} \textbf{2014}, \emph{118},
  4647--4660\relax
\mciteBstWouldAddEndPuncttrue
\mciteSetBstMidEndSepPunct{\mcitedefaultmidpunct}
{\mcitedefaultendpunct}{\mcitedefaultseppunct}\relax
\EndOfBibitem
\bibitem[Gao and Guo(2015)Gao, and Guo]{RN478}
Gao,~P.; Guo,~H. Developing coarse-grained potentials for the prediction of
  multi-properties of trans-1,4-polybutadiene melt. \emph{Polymer}
  \textbf{2015}, \emph{69}, 25--38\relax
\mciteBstWouldAddEndPuncttrue
\mciteSetBstMidEndSepPunct{\mcitedefaultmidpunct}
{\mcitedefaultendpunct}{\mcitedefaultseppunct}\relax
\EndOfBibitem
\bibitem[Sauter and Grafmüller(2017)Sauter, and Grafmüller]{RN222}
Sauter,~J.; Grafmüller,~A. Procedure for Transferable Coarse-Grained Models of
  Aqueous Polysaccharides. \emph{J. Chem. Theory Comput.} \textbf{2017},
  \emph{13}, 223--236\relax
\mciteBstWouldAddEndPuncttrue
\mciteSetBstMidEndSepPunct{\mcitedefaultmidpunct}
{\mcitedefaultendpunct}{\mcitedefaultseppunct}\relax
\EndOfBibitem
\bibitem[Abbott and Stevens(2015)Abbott, and Stevens]{RN161}
Abbott,~L.~J.; Stevens,~M.~J. A temperature-dependent coarse-grained model for
  the thermoresponsive polymer poly(N-isopropylacrylamide). \emph{J. Chem.
  Phys.} \textbf{2015}, \emph{143}\relax
\mciteBstWouldAddEndPuncttrue
\mciteSetBstMidEndSepPunct{\mcitedefaultmidpunct}
{\mcitedefaultendpunct}{\mcitedefaultseppunct}\relax
\EndOfBibitem
\bibitem[Kmiecik \latin{et~al.}(2016)Kmiecik, Gront, Kolinski, Wieteska, Dawid,
  and Kolinski]{RN4}
Kmiecik,~S.; Gront,~D.; Kolinski,~M.; Wieteska,~L.; Dawid,~A.~E.; Kolinski,~A.
  Coarse-Grained Protein Models and Their Applications. \emph{Chem. Rev.}
  \textbf{2016}, \relax
\mciteBstWouldAddEndPunctfalse
\mciteSetBstMidEndSepPunct{\mcitedefaultmidpunct}
{}{\mcitedefaultseppunct}\relax
\EndOfBibitem
\bibitem[Poma \latin{et~al.}(2017)Poma, Cieplak, and Theodorakis]{RN252}
Poma,~A.~B.; Cieplak,~M.; Theodorakis,~P.~E. Combining the MARTINI and
  Structure-Based Coarse-Grained Approaches for the Molecular Dynamics Studies
  of Conformational Transitions in Proteins. \emph{J. Chem. Theory Comput.}
  \textbf{2017}, \emph{13}, 1366--1374\relax
\mciteBstWouldAddEndPuncttrue
\mciteSetBstMidEndSepPunct{\mcitedefaultmidpunct}
{\mcitedefaultendpunct}{\mcitedefaultseppunct}\relax
\EndOfBibitem
\bibitem[Cheon \latin{et~al.}(2010)Cheon, Chang, and Hall]{RN1000}
Cheon,~M.; Chang,~I.; Hall,~C.~K. Extending the PRIME model for protein
  aggregation to all 20 amino acids. \emph{Proteins: Struct., Funct., Bioinf.}
  \textbf{2010}, \emph{78}, 2950--2960\relax
\mciteBstWouldAddEndPuncttrue
\mciteSetBstMidEndSepPunct{\mcitedefaultmidpunct}
{\mcitedefaultendpunct}{\mcitedefaultseppunct}\relax
\EndOfBibitem
\bibitem[Gopal~Srinivasa \latin{et~al.}(2009)Gopal~Srinivasa, Mukherjee, Cheng,
  and Feig]{RN652}
Gopal~Srinivasa,~M.; Mukherjee,~S.; Cheng,~Y.; Feig,~M. PRIMO/PRIMONA: A
  coarse‐grained model for proteins and nucleic acids that preserves
  near‐atomistic accuracy. \emph{Proteins: Struct., Funct., Bioinf.}
  \textbf{2009}, \emph{78}, 1266--1281\relax
\mciteBstWouldAddEndPuncttrue
\mciteSetBstMidEndSepPunct{\mcitedefaultmidpunct}
{\mcitedefaultendpunct}{\mcitedefaultseppunct}\relax
\EndOfBibitem
\bibitem[Bereau and Deserno(2009)Bereau, and Deserno]{RN650}
Bereau,~T.; Deserno,~M. Generic coarse-grained model for protein folding and
  aggregation. \emph{J. Chem. Phys.} \textbf{2009}, \emph{130}, 235106\relax
\mciteBstWouldAddEndPuncttrue
\mciteSetBstMidEndSepPunct{\mcitedefaultmidpunct}
{\mcitedefaultendpunct}{\mcitedefaultseppunct}\relax
\EndOfBibitem
\bibitem[Haxton \latin{et~al.}(2015)Haxton, Mannige, Zuckermann, and
  Whitelam]{RN283}
Haxton,~T.~K.; Mannige,~R.~V.; Zuckermann,~R.~N.; Whitelam,~S. Modeling
  Sequence-Specific Polymers Using Anisotropic Coarse-Grained Sites Allows
  Quantitative Comparison with Experiment. \emph{J. Chem. Theory Comput.}
  \textbf{2015}, \emph{11}, 303--315\relax
\mciteBstWouldAddEndPuncttrue
\mciteSetBstMidEndSepPunct{\mcitedefaultmidpunct}
{\mcitedefaultendpunct}{\mcitedefaultseppunct}\relax
\EndOfBibitem
\bibitem[Marrink \latin{et~al.}(2007)Marrink, Risselada, Yefimov, Tieleman, and
  de~Vries]{RN274}
Marrink,~S.~J.; Risselada,~H.~J.; Yefimov,~S.; Tieleman,~D.~P.; de~Vries,~A.~H.
  The MARTINI Force Field: Coarse Grained Model for Biomolecular Simulations.
  \emph{J. Phys. Chem. B} \textbf{2007}, \emph{111}, 7812--7824\relax
\mciteBstWouldAddEndPuncttrue
\mciteSetBstMidEndSepPunct{\mcitedefaultmidpunct}
{\mcitedefaultendpunct}{\mcitedefaultseppunct}\relax
\EndOfBibitem
\bibitem[Cornell \latin{et~al.}(1995)Cornell, Cieplak, Bayly, Gould, Merz,
  Ferguson, Spellmeyer, Fox, Caldwell, and Kollman]{RN681}
Cornell,~W.~D.; Cieplak,~P.; Bayly,~C.~I.; Gould,~I.~R.; Merz,~K.~M.;
  Ferguson,~D.~M.; Spellmeyer,~D.~C.; Fox,~T.; Caldwell,~J.~W.; Kollman,~P.~A.
  A Second Generation Force Field for the Simulation of Proteins, Nucleic
  Acids, and Organic Molecules. \emph{J. Am. Chem. Soc.} \textbf{1995},
  \emph{117}, 5179--5197\relax
\mciteBstWouldAddEndPuncttrue
\mciteSetBstMidEndSepPunct{\mcitedefaultmidpunct}
{\mcitedefaultendpunct}{\mcitedefaultseppunct}\relax
\EndOfBibitem
\bibitem[Jorgensen \latin{et~al.}(1996)Jorgensen, Maxwell, and
  Tirado-Rives]{RN682}
Jorgensen,~W.~L.; Maxwell,~D.~S.; Tirado-Rives,~J. Development and Testing of
  the OPLS All-Atom Force Field on Conformational Energetics and Properties of
  Organic Liquids. \emph{J. Am. Chem. Soc.} \textbf{1996}, \emph{118},
  11225--11236\relax
\mciteBstWouldAddEndPuncttrue
\mciteSetBstMidEndSepPunct{\mcitedefaultmidpunct}
{\mcitedefaultendpunct}{\mcitedefaultseppunct}\relax
\EndOfBibitem
\bibitem[Prakash \latin{et~al.}(2018)Prakash, Baer, Mundy, and
  Pfaendtner]{RN648}
Prakash,~A.; Baer,~M.~D.; Mundy,~C.~J.; Pfaendtner,~J. Peptoid Backbone
  Flexibilility Dictates Its Interaction with Water and Surfaces: A Molecular
  Dynamics Investigation. \emph{Biomacromolecules} \textbf{2018}, \emph{19},
  1006--1015\relax
\mciteBstWouldAddEndPuncttrue
\mciteSetBstMidEndSepPunct{\mcitedefaultmidpunct}
{\mcitedefaultendpunct}{\mcitedefaultseppunct}\relax
\EndOfBibitem
\bibitem[Wolfgang \latin{et~al.}(2011)Wolfgang, Thomas, and Ludger]{RN660}
Wolfgang,~B.; Thomas,~H.; Ludger,~W. Systematic conformational investigations
  of peptoids and peptoid–peptide chimeras. \emph{J. Pept. Sci.}
  \textbf{2011}, \emph{96}, 651--668\relax
\mciteBstWouldAddEndPuncttrue
\mciteSetBstMidEndSepPunct{\mcitedefaultmidpunct}
{\mcitedefaultendpunct}{\mcitedefaultseppunct}\relax
\EndOfBibitem
\bibitem[Mirijanian \latin{et~al.}(2014)Mirijanian, Mannige, Zuckermann, and
  Whitelam]{RN17}
Mirijanian,~D.~T.; Mannige,~R.~V.; Zuckermann,~R.~N.; Whitelam,~S. Development
  and use of an atomistic CHARMM-based forcefield for peptoid simulation.
  \emph{J Comput. Chem.} \textbf{2014}, \emph{35}, 360--70\relax
\mciteBstWouldAddEndPuncttrue
\mciteSetBstMidEndSepPunct{\mcitedefaultmidpunct}
{\mcitedefaultendpunct}{\mcitedefaultseppunct}\relax
\EndOfBibitem
\bibitem[MacKerell \latin{et~al.}(1998)MacKerell, Bashford, Bellott, Dunbrack,
  Evanseck, Field, Fischer, Gao, Guo, Ha, Joseph-McCarthy, Kuchnir, Kuczera,
  Lau, Mattos, Michnick, Ngo, Nguyen, Prodhom, Reiher, Roux, Schlenkrich,
  Smith, Stote, Straub, Watanabe, Wiórkiewicz-Kuczera, Yin, and
  Karplus]{RN700}
MacKerell,~A.~D.; Bashford,~D.; Bellott,~M.; Dunbrack,~R.~L.; Evanseck,~J.~D.;
  Field,~M.~J.; Fischer,~S.; Gao,~J.; Guo,~H.; Ha,~S. \latin{et~al.}  All-Atom
  Empirical Potential for Molecular Modeling and Dynamics Studies of Proteins.
  \emph{J. Phys. Chem. B} \textbf{1998}, \emph{102}, 3586--3616\relax
\mciteBstWouldAddEndPuncttrue
\mciteSetBstMidEndSepPunct{\mcitedefaultmidpunct}
{\mcitedefaultendpunct}{\mcitedefaultseppunct}\relax
\EndOfBibitem
\bibitem[Jorgensen \latin{et~al.}(1983)Jorgensen, Chandrasekhar, Madura, Impey,
  and Klein]{RN1002}
Jorgensen,~W.~L.; Chandrasekhar,~J.; Madura,~J.~D.; Impey,~R.~W.; Klein,~M.~L.
  Comparison of simple potential functions for simulating liquid water.
  \emph{J. Chem. Phys.} \textbf{1983}, \emph{79}, 926--935\relax
\mciteBstWouldAddEndPuncttrue
\mciteSetBstMidEndSepPunct{\mcitedefaultmidpunct}
{\mcitedefaultendpunct}{\mcitedefaultseppunct}\relax
\EndOfBibitem
\bibitem[Hess \latin{et~al.}(1997)Hess, Bekker, Berendsen, and Fraaije]{RN1005}
Hess,~B.; Bekker,~H.; Berendsen,~H. J.~C.; Fraaije,~J. G. E.~M. LINCS: A linear
  constraint solver for molecular simulations. \emph{J. Comput. Chem.}
  \textbf{1997}, \emph{18}, 1463--1472\relax
\mciteBstWouldAddEndPuncttrue
\mciteSetBstMidEndSepPunct{\mcitedefaultmidpunct}
{\mcitedefaultendpunct}{\mcitedefaultseppunct}\relax
\EndOfBibitem
\bibitem[Essmann \latin{et~al.}(1995)Essmann, Perera, Berkowitz, Darden, Lee,
  and Pedersen]{RN699}
Essmann,~U.; Perera,~L.; Berkowitz,~M.~L.; Darden,~T.; Lee,~H.; Pedersen,~L.~G.
  A smooth particle mesh Ewald method. \emph{J. Chem. Phys.} \textbf{1995},
  \emph{103}, 8577--8593\relax
\mciteBstWouldAddEndPuncttrue
\mciteSetBstMidEndSepPunct{\mcitedefaultmidpunct}
{\mcitedefaultendpunct}{\mcitedefaultseppunct}\relax
\EndOfBibitem
\bibitem[Bussi \latin{et~al.}(2007)Bussi, Donadio, and Parrinello]{RN635}
Bussi,~G.; Donadio,~D.; Parrinello,~M. Canonical sampling through velocity
  rescaling. \emph{J. Chem. Phys.} \textbf{2007}, \emph{126}, 014101\relax
\mciteBstWouldAddEndPuncttrue
\mciteSetBstMidEndSepPunct{\mcitedefaultmidpunct}
{\mcitedefaultendpunct}{\mcitedefaultseppunct}\relax
\EndOfBibitem
\bibitem[Berendsen \latin{et~al.}(1984)Berendsen, Postma, van Gunsteren,
  DiNola, and Haak]{RN636}
Berendsen,~H. J.~C.; Postma,~J. P.~M.; van Gunsteren,~W.~F.; DiNola,~A.;
  Haak,~J.~R. Molecular dynamics with coupling to an external bath. \emph{J.
  Chem. Phys.} \textbf{1984}, \emph{81}, 3684--3690\relax
\mciteBstWouldAddEndPuncttrue
\mciteSetBstMidEndSepPunct{\mcitedefaultmidpunct}
{\mcitedefaultendpunct}{\mcitedefaultseppunct}\relax
\EndOfBibitem
\bibitem[Evans and Holian(1985)Evans, and Holian]{RN637}
Evans,~D.~J.; Holian,~B.~L. The Nose–Hoover thermostat. \emph{J. Chem. Phys.}
  \textbf{1985}, \emph{83}, 4069--4074\relax
\mciteBstWouldAddEndPuncttrue
\mciteSetBstMidEndSepPunct{\mcitedefaultmidpunct}
{\mcitedefaultendpunct}{\mcitedefaultseppunct}\relax
\EndOfBibitem
\bibitem[Parrinello and Rahman(1980)Parrinello, and Rahman]{RN638}
Parrinello,~M.; Rahman,~A. Crystal Structure and Pair Potentials: A
  Molecular-Dynamics Study. \emph{Phys. Rev. Lett.} \textbf{1980}, \emph{45},
  1196--1199\relax
\mciteBstWouldAddEndPuncttrue
\mciteSetBstMidEndSepPunct{\mcitedefaultmidpunct}
{\mcitedefaultendpunct}{\mcitedefaultseppunct}\relax
\EndOfBibitem
\bibitem[Abraham \latin{et~al.}(2015)Abraham, Murtola, Schulz, Páll, Smith,
  Hess, and Lindahl]{RN639}
Abraham,~M.~J.; Murtola,~T.; Schulz,~R.; Páll,~S.; Smith,~J.~C.; Hess,~B.;
  Lindahl,~E. GROMACS: High performance molecular simulations through
  multi-level parallelism from laptops to supercomputers. \emph{SoftwareX}
  \textbf{2015}, \emph{1-2}, 19--25\relax
\mciteBstWouldAddEndPuncttrue
\mciteSetBstMidEndSepPunct{\mcitedefaultmidpunct}
{\mcitedefaultendpunct}{\mcitedefaultseppunct}\relax
\EndOfBibitem
\bibitem[Humphrey \latin{et~al.}(1996)Humphrey, Dalke, and Schulten]{RN640}
Humphrey,~W.; Dalke,~A.; Schulten,~K. VMD: Visual molecular dynamics. \emph{J.
  Mol. Graph.} \textbf{1996}, \emph{14}, 33--38\relax
\mciteBstWouldAddEndPuncttrue
\mciteSetBstMidEndSepPunct{\mcitedefaultmidpunct}
{\mcitedefaultendpunct}{\mcitedefaultseppunct}\relax
\EndOfBibitem
\bibitem[Bennett(1976)]{RN1006}
Bennett,~C.~H. Efficient estimation of free energy differences from Monte Carlo
  data. \emph{J. Comput. Phys.} \textbf{1976}, \emph{22}, 245 -- 268\relax
\mciteBstWouldAddEndPuncttrue
\mciteSetBstMidEndSepPunct{\mcitedefaultmidpunct}
{\mcitedefaultendpunct}{\mcitedefaultseppunct}\relax
\EndOfBibitem
\bibitem[Kumar \latin{et~al.}(1992)Kumar, Rosenberg, Bouzida, Swendsen, and
  Kollman]{RN1007}
Kumar,~S.; Rosenberg,~J.~M.; Bouzida,~D.; Swendsen,~R.~H.; Kollman,~P.~A. THE
  weighted histogram analysis method for free-energy calculations on
  biomolecules. I. The method. \emph{J. Comput. Chem.} \textbf{1992},
  \emph{13}, 1011--1021\relax
\mciteBstWouldAddEndPuncttrue
\mciteSetBstMidEndSepPunct{\mcitedefaultmidpunct}
{\mcitedefaultendpunct}{\mcitedefaultseppunct}\relax
\EndOfBibitem
\bibitem[Jorge \latin{et~al.}(2010)Jorge, Garrido, Queimada, Economou, and
  Macedo]{RN1008}
Jorge,~M.; Garrido,~N.~M.; Queimada,~A.~J.; Economou,~I.~G.; Macedo,~E.~A.
  Effect of the Integration Method on the Accuracy and Computational Efficiency
  of Free Energy Calculations Using Thermodynamic Integration. \emph{J. Chem.
  Theory Comput.} \textbf{2010}, \emph{6}, 1018--1027\relax
\mciteBstWouldAddEndPuncttrue
\mciteSetBstMidEndSepPunct{\mcitedefaultmidpunct}
{\mcitedefaultendpunct}{\mcitedefaultseppunct}\relax
\EndOfBibitem
\bibitem[Taddese and Carbone(2017)Taddese, and Carbone]{RN284}
Taddese,~T.; Carbone,~P. Effect of Chain Length on the Partition Properties of
  Poly(ethylene oxide): Comparison between MARTINI Coarse-Grained and Atomistic
  Models. \emph{J. Phys. Chem. B} \textbf{2017}, \emph{121}, 1601--1609\relax
\mciteBstWouldAddEndPuncttrue
\mciteSetBstMidEndSepPunct{\mcitedefaultmidpunct}
{\mcitedefaultendpunct}{\mcitedefaultseppunct}\relax
\EndOfBibitem
\bibitem[Dalgicdir \latin{et~al.}(2013)Dalgicdir, Sensoy, Peter, and
  Sayar]{RN2}
Dalgicdir,~C.; Sensoy,~O.; Peter,~C.; Sayar,~M. A transferable coarse-grained
  model for diphenylalanine: how to represent an environment driven
  conformational transition. \emph{J. Chem. Phys.} \textbf{2013}, \emph{139},
  234115\relax
\mciteBstWouldAddEndPuncttrue
\mciteSetBstMidEndSepPunct{\mcitedefaultmidpunct}
{\mcitedefaultendpunct}{\mcitedefaultseppunct}\relax
\EndOfBibitem
\bibitem[Ryckaert and Bellemans(1978)Ryckaert, and Bellemans]{RN2000}
Ryckaert,~J.-P.; Bellemans,~A. Molecular dynamics of liquid alkanes.
  \emph{Faraday Discuss. Chem. Soc.} \textbf{1978}, \emph{66}, 95--106\relax
\mciteBstWouldAddEndPuncttrue
\mciteSetBstMidEndSepPunct{\mcitedefaultmidpunct}
{\mcitedefaultendpunct}{\mcitedefaultseppunct}\relax
\EndOfBibitem
\bibitem[Uttarwar \latin{et~al.}(2013)Uttarwar, Potoff, and Huang]{RN688}
Uttarwar,~R.~G.; Potoff,~J.; Huang,~Y. Study on Interfacial Interaction between
  Polymer and Nanoparticle in a Nanocoating Matrix: A MARTINI Coarse-Graining
  Method. \emph{Ind. Eng. Chem. Res.} \textbf{2013}, \emph{52}, 73--82\relax
\mciteBstWouldAddEndPuncttrue
\mciteSetBstMidEndSepPunct{\mcitedefaultmidpunct}
{\mcitedefaultendpunct}{\mcitedefaultseppunct}\relax
\EndOfBibitem
\bibitem[Bordner \latin{et~al.}(2002)Bordner, Cavasotto, and Abagyan]{RN508}
Bordner,~A.~J.; Cavasotto,~C.~N.; Abagyan,~R.~A. Accurate Transferable Model
  for Water, n-Octanol, and n-Hexadecane Solvation Free Energies. \emph{J.
  Phys. Chem. B} \textbf{2002}, \emph{106}, 11009--11015\relax
\mciteBstWouldAddEndPuncttrue
\mciteSetBstMidEndSepPunct{\mcitedefaultmidpunct}
{\mcitedefaultendpunct}{\mcitedefaultseppunct}\relax
\EndOfBibitem
\bibitem[Monticelli \latin{et~al.}(2008)Monticelli, Kandasamy, Periole, Larson,
  Tieleman, and Marrink]{RN22}
Monticelli,~L.; Kandasamy,~S.~K.; Periole,~X.; Larson,~R.~G.; Tieleman,~D.~P.;
  Marrink,~S.~J. The MARTINI coarse-grained force field: Extension to proteins.
  \emph{J. Chem. Theory Comput.} \textbf{2008}, \emph{4}, 819--834\relax
\mciteBstWouldAddEndPuncttrue
\mciteSetBstMidEndSepPunct{\mcitedefaultmidpunct}
{\mcitedefaultendpunct}{\mcitedefaultseppunct}\relax
\EndOfBibitem
\bibitem[Murnen \latin{et~al.}(2012)Murnen, Khokhlov, Khalatur, Segalman, and
  Zuckermann]{RN207}
Murnen,~H.~K.; Khokhlov,~A.~R.; Khalatur,~P.~G.; Segalman,~R.~A.;
  Zuckermann,~R.~N. Impact of Hydrophobic Sequence Patterning on the
  Coil-to-Globule Transition of Protein-like Polymers. \emph{Macromolecules}
  \textbf{2012}, \emph{45}, 5229--5236\relax
\mciteBstWouldAddEndPuncttrue
\mciteSetBstMidEndSepPunct{\mcitedefaultmidpunct}
{\mcitedefaultendpunct}{\mcitedefaultseppunct}\relax
\EndOfBibitem
\bibitem[Rubinstein and Colby(2003)Rubinstein, and Colby]{RN672}
Rubinstein,~M.; Colby,~R.~H. \emph{Polymer Physics}; Oxford University Press,
  2003\relax
\mciteBstWouldAddEndPuncttrue
\mciteSetBstMidEndSepPunct{\mcitedefaultmidpunct}
{\mcitedefaultendpunct}{\mcitedefaultseppunct}\relax
\EndOfBibitem
\bibitem[Tukhvatullin \latin{et~al.}(1999)Tukhvatullin, Tashkenbaev, Zhumaboev,
  and Mamatov]{RN680}
Tukhvatullin,~F.~K.; Tashkenbaev,~U.~N.; Zhumaboev,~A.; Mamatov,~Z.
  Intermolecular hydrogen bonds in acetic acid and its solutions. \emph{J.
  Appl. Spectrosc.} \textbf{1999}, \emph{66}, 501--505\relax
\mciteBstWouldAddEndPuncttrue
\mciteSetBstMidEndSepPunct{\mcitedefaultmidpunct}
{\mcitedefaultendpunct}{\mcitedefaultseppunct}\relax
\EndOfBibitem
\bibitem[Dalgicdir \latin{et~al.}(2016)Dalgicdir, Globisch, Sayar, and
  Peter]{RN481}
Dalgicdir,~C.; Globisch,~C.; Sayar,~M.; Peter,~C. Representing
  environment-induced helix-coil transitions in a coarse grained peptide model.
  \emph{Eur. Phys. J. Spec. Top.} \textbf{2016}, \emph{225}, 1463--1481\relax
\mciteBstWouldAddEndPuncttrue
\mciteSetBstMidEndSepPunct{\mcitedefaultmidpunct}
{\mcitedefaultendpunct}{\mcitedefaultseppunct}\relax
\EndOfBibitem
\bibitem[Shi \latin{et~al.}(2018)Shi, Wei, Zhu, Sun, and Li]{RN686}
Shi,~Z.; Wei,~Y.; Zhu,~C.; Sun,~J.; Li,~Z. Crystallization-Driven
  Two-Dimensional Nanosheet from Hierarchical Self-Assembly of
  Polypeptoid-Based Diblock Copolymers. \emph{Macromolecules} \textbf{2018},
  \emph{51}, 6344--6351\relax
\mciteBstWouldAddEndPuncttrue
\mciteSetBstMidEndSepPunct{\mcitedefaultmidpunct}
{\mcitedefaultendpunct}{\mcitedefaultseppunct}\relax
\EndOfBibitem
\end{mcitethebibliography}

% I'm adding this graphical TOC entry as-is for now, but advise that it might need explanatory labels of some kind. --Andrew
\begin{tocentry}
	\includegraphics[scale=0.8]{media/1_2.png}
\end{tocentry}

\end{document}